\documentclass[10pt,amsmath,amssymb,aps, prl, twocolumn,floatfix,superscriptaddress,preprintnumbers,showpacs]{revtex4-2}
\usepackage[T1]{fontenc}
\usepackage[utf8]{inputenc}
\usepackage{xcolor}
\usepackage{babel}
\usepackage{verbatim}
\usepackage{graphicx}
\usepackage[colorlinks=true,
			linkcolor=magenta,
			citecolor=blue,
			urlcolor=black,]{hyperref}
\usepackage[caption=false,labelfont=bf]{subfig}
\usepackage{float}

\begin{document}
\title{QSSEP describes the fluctuations of quantum coherences in the Anderson model}
\author{Ludwig Hruza}
\affiliation{LPENS, Département de physique, École normale supérieure, Université PSL, Sorbonne Université, Université Paris Cité, CNRS, 75005 Paris.}
\affiliation{Laboratoire de Neurosciences Cognitives et Computationnelles, École normale supérieure, Université PSL, INSERM, 75005 Paris, France}
\author{Tony Jin}
\affiliation{Universit\'e C\^ote-d'Azur, CNRS, Centrale Med, Institut de Physique de Nice, 06200 Nice, France}
\begin{abstract}
Using the transfer matrix method, we numerically investigate the
structure of spatial coherences and their fluctuations in the 3d
Anderson model in the metallic phase when driven out-of-equilibrium
by external leads at zero temperature and in linear response. We find
that the stationary state entails non-local non-Gaussian correlations
in the longitudinal direction, which are characteristic of diffusive
non equilibrium steady states. These correlations are \emph{quantitatively} matched, at least up to third order, by those analytically derived in the Quantum Symmetric Simple Exclusion Process (QSSEP)
which describes diffusive fermions in 1d subject to \emph{dynamical} disorder. 
Furthermore, the large deviation scaling and $U(1)$ invariance of these correlations imply a link between the Anderson model and free probability theory.
Our findings suggest the existence of a universal structure of correlations in non-interacting diffusive quantum systems that might be captured by QSSEP.
\end{abstract}
 \maketitle
In a seminal paper in 1995, Lee, Levitov and Yakovets \citep{LeeLevitovYuPRBUniversalstatistics}
computed the generating function of the integrated current $Q_{t}$
up to time $t$ passing through a quasi-1d metallic wire, using a
Landauer approach at low temperature and in linear response. A few
years later, in 2003, Derrida, Douçot and Roche \citep{RocheDerridaCurrentfluctuationsinSSEP}
managed to perform the same derivation on an entirely different model,
the Symmetric Simple Exclusion Process (SSEP) \citep{Derrida_ReviewSSEP}. Much to their surprise,
they found the \emph{exact same result}. To paraphrase them, this
is anything but a trivial connection, since Lee at al.\ studied diffusive wires with \emph{static disorder} on the basis of the Dorokhov-Mello-Pereyra-Kumar (DMPK) formalism \citep{DMPK1,DMPK2}, while the computation from Derrida et al.\ was done on a
purely classical model with \emph{dynamical }disorder. 

\begin{figure}
\begin{centering}
\includegraphics[width=0.92\columnwidth]{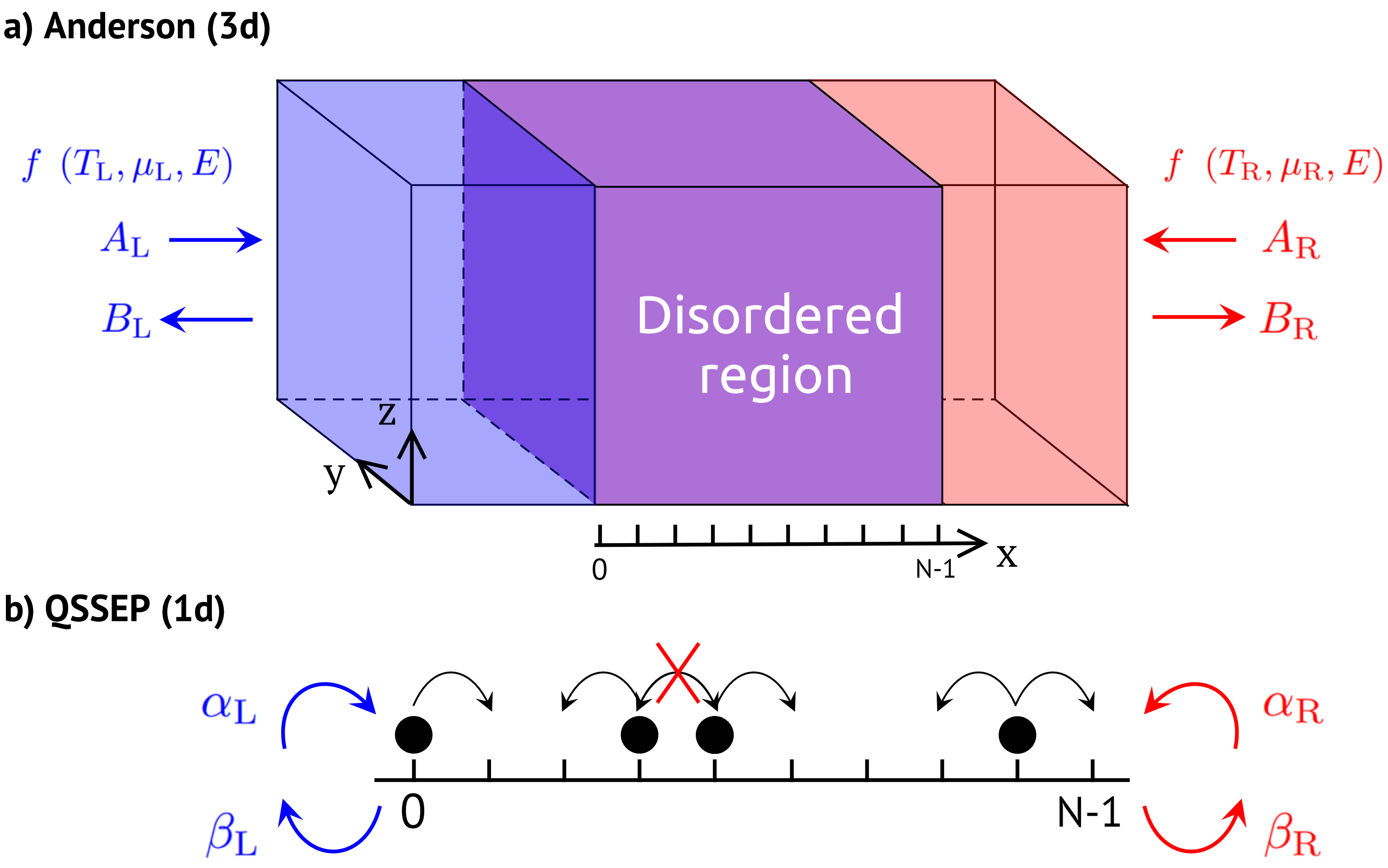}
\par\end{centering}
\caption{The two models investigated in this paper. \textbf{a)} The Anderson model
in 3d. Electrons propagate from the leads (clean regions)
denoted left/right
(${\rm L/R}$) through a region with \emph{static }disordered potential
in purple. The leads are taken at thermal equilibrium with Fermi
distributions $f(T_{{\rm L/R}},\mu_{{\rm L/R}},E)$. Here
$A_{{\rm R/L}}$ denotes the incoming modes and $B_{{\rm R/L}}$ the
outgoing ones. \textbf{b)} The Quantum Symmetric Simple Exclusion
Process (QSSEP) in 1d. The system is driven out-of-equilibrium by Lindblad
boundary terms injecting and extracting
particles at respective rates $\alpha_{{\rm L/R}}$, $\beta_{{\rm L/R}}$.
The bulk dynamics is given by Eq.~(\ref{eq:Q-SSEPHamiltonian}) and models
fermions undergoing coherent jumps with random amplitudes both in time \emph{and} space. \label{fig:The-two-models}}
\end{figure}

Not long ago, an extension of SSEP to the quantum realm named QSSEP
was proposed and studied in \citep{BauerBernardJin_EquilibriumQSSEP,BernardJin_QSSEP}.
The QSSEP reproduces in mean the density and current fluctuations of the SSEP (similarly to related models \citep{Eisler_CrossoverBallisticDiffusive,VerstraeteSEP})
but additionally entails fluctuations of non-classical quantities
such as quantum coherence, see Eq. (\ref{eq:gQSSEP-1}), or entanglement.
In this paper, we wish to explore further this intriguing connection
by comparing the fluctuations of quantum coherence in the
3d Anderson model in the metallic regime to known
analytical results from QSSEP. Another motivation to study the 3d Anderson
model
is that its mutual information, investigated numerically in \citep{GullansHuseEntanglementstructure},
was shown to be extensive which is also the case in QSSEP as a recent
exact result demonstrates \citep{HruzaBenardEntanglementQSSEP}.

Surprisingly, we find that QSSEP indeed describes \emph{quantitatively
}the fluctuations of spatial coherences in the Anderson model, at least up to third order. In
particular, both models entail the presence of \emph{\mbox{non-local} \mbox{non-Gaussian}} correlations which are characteristic of the non-equilibrium steady
state (NESS) of diffusive systems. This brings forth the exciting
possibility that QSSEP provides an effective description of diffusive
metallic wires. As the statistics of QSSEP was recently connected
to free probability theory \citep{Biane2021Combinatorics,HruzaBernardPRX}, our finding suggests the existence of a hidden free probability structure
in the NESS of diffusive metallic wires \footnote{An obvious but important
point to mention is that this mapping, if it were to exist, can not
hold in a generic matter. For instance, it is known that, contrarily
to the Anderson model, there is no localized phase for the QSSEP model.
We thus restrict our conjecture to the metallic phase of the
3d Anderson model.}.

Both in a 3d and quasi-1d geometry, the shape of the spatial correlations in the Anderson model is accurately captured by QSSEP. However the correct scaling with system size only matches in 3d. We leave the interesting discussion of the quasi-1d case to the SM \cite{SM}.

We begin by presenting the models and recall
some useful results and methodology. We then present our numerical
results concerning the Anderson model and show their quantitative
agreements with the exact known results from QSSEP. Finally, we propose
some elements and a roadmap for proving the correspondence analytically.
We conclude by discussing exciting open questions. 

\paragraph{Anderson model \citep{AndersonOGpaper}.} We consider the Hamiltonian
\begin{align}
H_{{\rm A}} & :=-\sum_{(\boldsymbol{i},\boldsymbol{j})\text{ neighbors}}\left(t_{(\boldsymbol{i},\boldsymbol{j})}c_{\boldsymbol{i}}^{\dagger}c_{\boldsymbol{j}}+{\rm h.c}\right)+\sum_{\boldsymbol{i}}V_{\boldsymbol{i}}c_{\boldsymbol{i}}^{\dagger}c_{\boldsymbol{i}}\label{eq:Anderson-1}
\end{align}
where $\boldsymbol{i}:=\left(i_{x},i_{y},i_{z}\right)$, $t_{(\boldsymbol{i},\boldsymbol{j})}=t_{x/y/z}$
for a link in the $x/y/z$ direction and $c_{\boldsymbol{i}}$ is the
usual fermionic annihilation operator, see Fig.~\ref{fig:The-two-models} a. 
The random onsite-potentials $V_{\boldsymbol{i}}$ are i.i.d.\ and uniform in $\left[-\frac{W}{2},\frac{W}{2}\right]$ inside a cubic disordered region of size $N^{3}$. On the boundaries in $x$-direction ($i_x=0,N-1$), we attach clean infinite leads ($W=0$) to the system which are described by Fermi distributions $f(T_{{\rm \alpha}},\mu_{\alpha},E)$, where $\alpha={\rm L,R}$ denotes the left and right lead.
The dispersion relation in the leads is given by $E_{\boldsymbol{k}}=-\sum_{\nu=x,y,z}2t_{\nu}\cos\left(k_{\nu}a\right)$
with $\boldsymbol{k}:=\left(k_{x},\frac{2\pi n}{Na},\frac{2\pi m}{Na}\right)$,
$n,m\text{ integers}\in[0,N-1],$ $a$ the lattice spacing and $k_{x}\in[-\frac{\pi}{a},\frac{\pi}{a}]$
a continuous index, as the total system is infinite in the $x$-direction. Note that fixing $E_{\boldsymbol{k}}$ and
transverse momenta $k_\perp:=(k_y,k_z)$ 
fixes $k_x(E,k_\perp)$. On each side, we denote $A_{\alpha}(E,k_\perp)$
the amplitudes of the incoming and $B_{\alpha}(E,k_\perp)$
the amplitudes of the outgoing plane waves with given energy and transverse momentum.
Since the system is non-interacting, we can describe the NESS using Landauer-Büttiker's formalism \citep{LandauerFormula,Buttikerformula}. 

This model has a well-known phase transition from
a metallic to a localized phase as $W$ is increased above a critical
value $W_{c}$ \citep{Markosnumerics__2006}. We will work in the metallic
phase $W_{c}<W$. Additionally, we are interested in the regime where
the mean-free path $\ell$ is small compared to the total system size
$\ell\ll Na$. 


\paragraph{QSSEP.} The Quantum Symmetric Simple Exclusion Process was introduced in \citep{BauerBernardJin_Stoqdissipative}
and subsequently studied in \citep{BauerBernardJin_EquilibriumQSSEP,BernardJin_QSSEP,BernardJin_SolutionQSSEPcontinue,Biane2021Combinatorics,BernardPiroli_QSSEPentanglement,LudwigDenisMarkoFabian_DynamicsQSSEP,HruzaBenardEntanglementQSSEP,HruzaBernardPRX}.
The QSSEP describes random hopping of fermionic particles on a discrete
1d lattice coupled to boundary reservoirs on each side. This model is explicitly diffusive, i.e.\ it fulfills Fick's law at the operatorial level. Remarkably, it is possible to obtain
all the spatial connected correlations for the out-of-equilibrium
stationary state \emph{at any order }in this model. The bulk Hamiltonian
is given by (see Fig.~\ref{fig:The-two-models} b)

\begin{align}
dH_{{\rm QSSEP}}(t) & :=\sqrt{D}\sum_{i=0}^{N-2}c_{i}^{\dagger}c_{i+1}dW_{t}^{i}+{\rm h.c}\label{eq:Q-SSEPHamiltonian}
\end{align}
where $W_{t}^{i}:=\frac{B_{t}^{1,i}+iB_{t}^{2,i}}{\sqrt{2}}$
with $B_{t}^{1,i}$ and $B_{t}^{2,i}$ independent Brownian processes with variance $\delta_{ij} t$, and $D$ the diffusion constant. \textcolor{black}{The density matrix evolves according
to $\rho_{t+dt}=e^{-idH_{t}}\rho_{t}e^{idH_{t}}.$} Additionally,
the system is driven out-of-equilibrium by Lindblad creation and annhilation
superoperators \textcolor{black}{${\cal D}[c^{\dagger}/c]$} acting
on the boundaries with rates $\alpha_{{\rm L/R}}$ for the creation
and $\beta_{{\rm L/R}}$ for the annihilation. For simplification,
we will fix $\alpha_{{\rm L/R}}+\beta_{{\rm L/R}}=1$ throughout the
manuscript. In the absence of bulk dynamics, the effect of the boundary
terms is to fix the density of the first site to $\alpha_{{\rm L}}$
and of the last site to $\alpha_{{\rm R}}$. 

There are $3$ types of disorder we will consider in the problem.
In what follows, we denote quantum averages in both models by $\langle\bullet\rangle$,
static disorder average in Anderson with $\overline{\bullet}$,
and average with respect to the fluctuating complex noise in QSSEP
by $\mathbb{E}[\bullet]$.

\paragraph{QSSEP correlations.}

We recall here relevant results on the spatial coherences
of QSSEP in the NESS \footnote{It is also possible to make statements about the dynamics of QSSEP \cite{HruzaBernardPRX}, but here we restrict ourselves to the steady state}. We are interested
in the two-point function
\begin{equation}
G_{ij}:={\rm tr}\left(\rho_{t=\infty}c_{j}^{\dagger}c_{i}\right)\label{eq:gQSSEP-1}
\end{equation}
and its fluctuations with respect to the noise. Because the dynamics preserves Gaussian fermionic states for each individual stochastic
trajectory, the statistics of QSSEP is fully encoded in $G_{ij}$. The cumulants of $G_{ij}$ satisfy a $U(1)$ invariance,
i.e. $\mathbb{E}[G_{i_{1}j_{1}}\cdots G_{i_{n}j_{n}}]^{c}$ is non-zero only if $\left\{ j_{1},\cdots j_{n}\right\} $ is a permutation of $\left\{ i_{1},\cdots i_{n}\right\}$.
Of all possible allowed cumulants, so-called ``cyclic cumulants'' $\mathbb{E}[G_{i_{1}i_{2}}G_{i_{2}i_{3}}\cdots G_{i_{n}i_{1}}]^{c}$
are of special importance: They scale as $N^{1-n}$ with the number
of sites $N$ and they constitute the leading order cumulant at any
order $n$. In the limit $N\to\infty$, with coordinates
$x_{k}:=\frac{i_{k}}{N}\in[0,1]$, we define
\begin{equation}
g_{n}^{{\rm Q}}\left(x_{1},\cdots,x_{n}\right):=\lim_{N\to\infty}N^{n-1}\mathbb{E}\left[G_{i_{1}i_{2}}G_{i_{2}i_{3}}\cdots G_{i_{n}i_{1}}\right]^{c}
\end{equation}
with $\Delta n=\alpha_{R}-\alpha_{L}$ the boundary imbalance. These
functions contain the complete information about QSSEP at leading order
in $N$. The first three orders are
\begin{align}
 & g_{1}^{{\rm Q}}=\Delta n\,x,\quad g_{2}^{{\rm Q}}=(\Delta n)^{2}\left(\min(x,y)-xy\right),\label{eq:QSSEPcontinuous-1}\\
 & g_{3}^{{\rm Q}}=(\Delta n)^{3}\left(\min(x,y,z)-(x\min(y,z))_{\circlearrowleft3}+2xyz\right)\nonumber 
\end{align}
and $(\cdots)_{\circlearrowleft3}$ denotes a sum of three terms
obtain by cyclic permutation of $(x,y,z)$. Note that these quantities are completely independent of the diffusion constant $D$.

Thanks to an elegant connection between
QSSEP and free probability 
\citep{Biane2021Combinatorics,HruzaBernardPRX},
the general solution can be formulated recursively as 
\begin{equation}
\sum_{\pi\in \rm{NC}(n)}\prod_{p\in \pi}g_{|p|}^{{\rm Q}}(\vec{x}_{p})=(\Delta n)^{n}\min(x_{1},\cdots,x_{n})
\end{equation}
with $\pi\in \rm{NC}(n)$ a non-crossing set-partition of $n$ elements,
$p$ the parts of this partition, $|p|$ the number of elements in
part $p$ and $\vec{x}_{p}=(x_{i})_{i\in p}$.

\paragraph{Numerical results for the Anderson model.}

We now present numerical results obtained in the 3d Anderson model
and will compare them with the exact results for the QSSEP. Our approach
relies on the Landauer-Büttiker formalism \citep{LandauerFormula,LandauerButtiker}
and the transfer matrix method for non-interacting fermionic systems \citep{MacKinnonKramerNumericsAnderson,Pendrynumerics,Markosnumerics__2006}
which we recall in the SM \citep{SM}. The numerics allows us to access single-particle 
eigenstates corresponding to particles incoming
from lead \mbox{$\alpha={\rm L,R}$} with energy $E$ and transverse momentum $k_\perp$
which we denote $\left|\psi_{\alpha,E,k_\perp}\right\rangle$. Let $a_{{\alpha,E,k_\perp}}^{\dagger}$
be the second quantized fermionic creation operator associated to $\left|\psi_{\alpha,E,k_\perp}\right\rangle $.
We fix the statistics of the bath by $\langle a_{\alpha,E,k_\perp}^{\dagger}a_{\alpha',E',k'_\perp}\rangle=\delta_{\alpha\alpha'}\delta\left(E,E'\right)\delta_{k_{\perp},k'_{\perp}}f\left(T_{\alpha},\mu_{\alpha},E\right)$. The two-point function $G_{\boldsymbol{ij}}^{{\rm A}}:=\langle c_{\boldsymbol{j}}^{\dagger}c_{\boldsymbol{i}}\rangle$ whose statistics we are interested in is
\begin{equation}
G_{\boldsymbol{ij}}^{{\rm A}}=\int dE\sum_{\alpha={\rm L,R}}f\left(T_{\alpha},\mu_{\alpha},E\right)\sum_{k_{\perp}}\psi_{\alpha,E,k_{\perp}}^{*}(\boldsymbol{j})\psi_{\alpha,E,k_{\perp}}(\boldsymbol{i}). \nonumber
\end{equation}
Imposing a small imbalance $\mu_{{\rm L/R}}=\mp\delta\mu$ between the leads
and fixing $T_{{\rm L}}=T_{{\rm R}}=0$, we can expand around $E=0$. Replacing the Fermi distributions $f$ by step functions and denoting $G_{\boldsymbol{ij}}^{\alpha}(E):=\sum_{k_{\perp}}\psi_{\alpha,E,k_{\perp}}^{*}(\boldsymbol{j})\psi_{\alpha, E,k_{\perp}}(\boldsymbol{i})$,
\begin{equation}
G^{{\rm A}}\approx\delta\mu(G^{{\rm R}}(0^{+})-G^{{\rm L}}(0^{-}))+\int_{-\infty}^{0}dE(G^{{\rm R}}(E)+G^{{\rm L}}(E)) \nonumber
\end{equation}
 and one identifies the non-equilibrium part of $G^\text{A}$ as
\begin{equation}
G^{\mathrm{neq}}:=\delta\mu\left(G^{{\rm R}}(0^{+})-G^{{\rm L}}(0^{-})\right),
\end{equation}
where $0^{\pm}$ means that we can evaluate the energy for any point
in the interval $(0,\pm\delta\mu]$. In our simulations we take $0^\pm=\pm0.2$. For convenience, we will avoid
dealing with imaginary values of $k_{x}$ by imposing an anisotropic
tight-binding term with $t_{y},t_{z}<\frac{1}{2}t_{x}$. Here we choose $t_{x}=1$ and $t_{y}=t_{z}=0.4$. We conventionally
fix the rate of incoming particle of a given mode to $1$ and $\delta\mu=1$. We fix the disorder strength $W$ (in the metallic regime) a posteriori to best fit the QSSEP prediction. At second order, we will see that this is the case for $W\approx6$. Other values of $W$ for which the agreement is worse are shown in the SM \citep{SM}.

\begin{figure}
    \centering
    \includegraphics[width=1\linewidth]{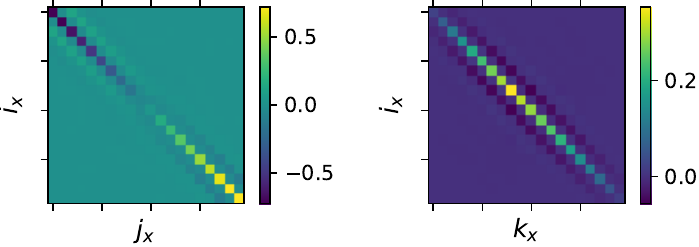}
    \vspace{-2mm}
    \caption{Numerical verification of the $U(1)$ invariance of cumululants in the Anderson model for $W=6$ and $N=20$. Left: $\left[\overline{G^\text{neq}_{\boldsymbol{i}\boldsymbol{j}}}\right]_\perp$ (with $i_\perp=j_\perp$) is nonzero only if $i_x=j_x$. Right: $N\left[\overline{G^\text{neq}_{\boldsymbol{i}\boldsymbol{j}}G^\text{neq}_{\boldsymbol{j}\boldsymbol{k}}}^c\right]_\perp$ (with $i_\perp=k_\perp$ and $j_x=8$) requires $i_x=k_x$.}
    \label{fig:U1_invariance}
\end{figure}

We will now define correlations functions $g_{n}^{A}$
for the Anderson model in analogy to $g_{n}^{{\rm Q}}$
for QSSEP and numerically investigate to what extent they agree, i.e.
\begin{equation}
\text{\ensuremath{g_{n}^{{\rm A}}\overset{?}{=}g_{n}^{{\rm Q}}}.}
\label{eq:conjecture}
\end{equation}
where the imbalance $\Delta n$ in the definition of $g_{n}^{{\rm Q}}$
is taken as a fitting parameter. Our proposal is to define
\begin{equation}\label{eq:eq:anderson_cumulants}
    g_{n}^{{\rm A}}(x_{1},\cdots,x_{n}):=N^{n-1}\left[\overline{G_{\boldsymbol{i}_{1}\boldsymbol{i}_{2}}^{{\rm neq}}\cdots G_{\boldsymbol{i}_{n}\boldsymbol{i}_{1}}^{{\rm neq}}}^{c}\right]_{\perp}
\end{equation}
with $x_{k}=i_{k,x}/N$ the rescaled positions. Here $\overline{\bullet}^{c}$
denotes the connected disorder average and $[\bullet]_{\perp}$denotes
the spatial average over perpendicular indices, which reduces the formerly
3d expression to 1d
\begin{equation}\label{eq:spatial_average}
\left[G_{\boldsymbol{i}_{1}\boldsymbol{i}_{2}}^{{\rm neq}}\cdots G_{\boldsymbol{i}_{n}\boldsymbol{i}_{1}}^{{\rm neq}}\right]_{\perp}:=\frac{1}{N^{2n}}\sum_{i_{1\perp},\cdots, i_{n\perp}}G_{\boldsymbol{i}_{1}\boldsymbol{i}_{2}}^{{\rm neq}}\cdots G_{\boldsymbol{i}_{n}\boldsymbol{i}_{1}}^{{\rm neq}}
\end{equation}
with $i_\perp = (i_y,i_z).$ Note that \eqref{eq:eq:anderson_cumulants} is $U(1)$ invariant by construction. We numerically verify in Fig.~\ref{fig:U1_invariance} that non-vanishing cumulants of coherences in the Anderson model must be $U(1)$ invariant. Importantly, this $U(1)$ invariance, together with the large deviation scaling \eqref{eq:eq:anderson_cumulants}, implies a link between the Anderson model and free probability theory \cite{HruzaBernardPRX}.

We conduct investigation of Eq.~\eqref{eq:conjecture} up to order $n=3$. 

\begin{figure*}
\centering{\includegraphics[width=0.95\textwidth]{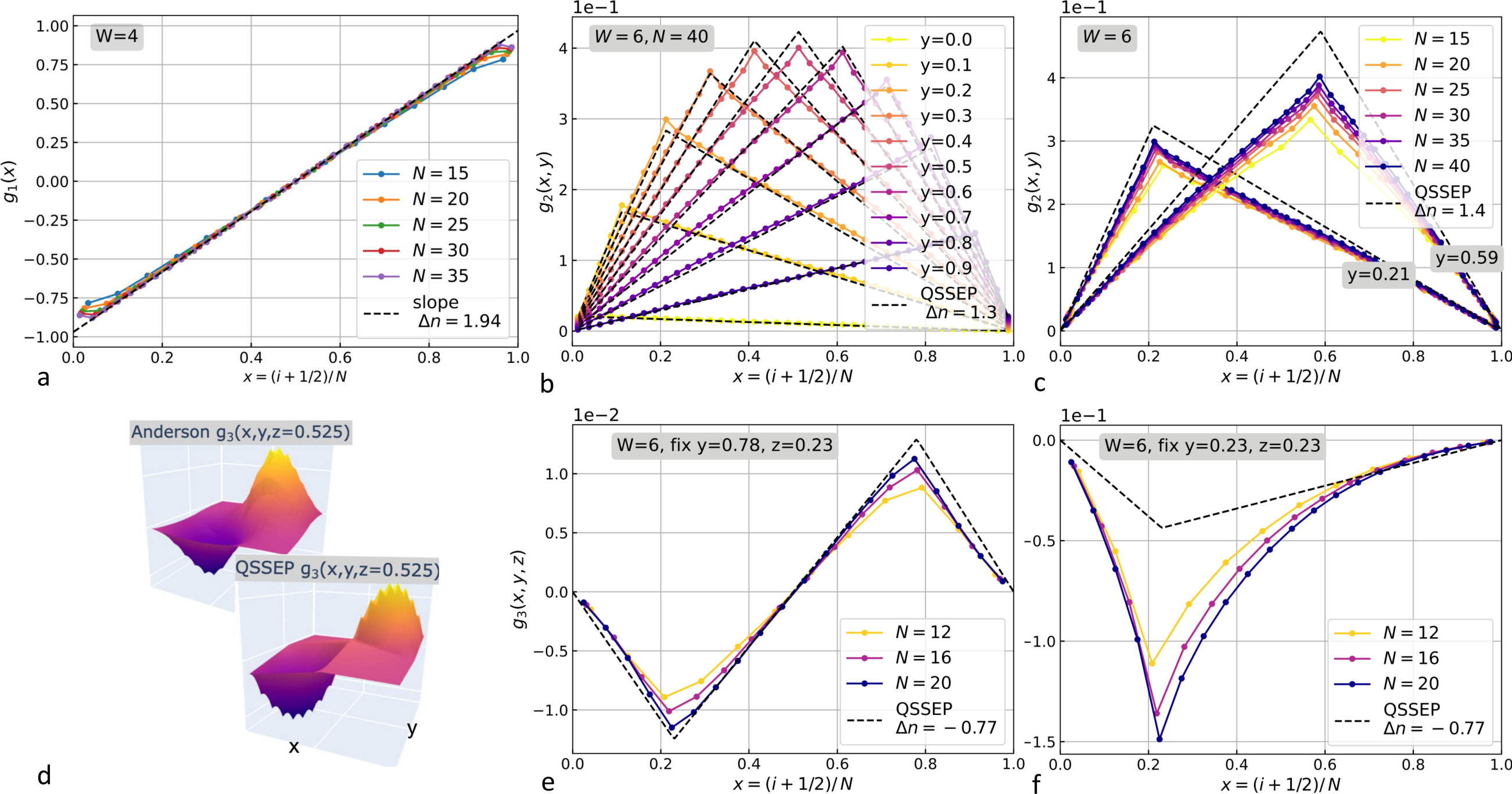}}
\caption{On all the figures, the dashed line corresponds to analytical predictions from QSSEP (\ref{eq:QSSEPcontinuous-1}) while the solid colored lines correspond to numerical simulations of the 3d Anderson model 
for different linear sizes $N$ and disorder $W=6$. We use the continuous notations for coordinates in $x$-direction, $x=\frac{i_x}{N}, y=\frac{j_x}{N}$ and $z:=\frac{k_x}{N}$. 
The disorder averages are performed over $1000$ realizations for every point. 
\textbf{a)}
The density $g^{{\rm Q/A}}_1$ is linear as expected for diffusive systems. Its slope is $\Delta n^{(1)}=1.75$.
\textbf{ b)} Comparison of the second cumulant $g^{{\rm Q/A}}_2$ between QSSEP and Anderson for $N=40$. The different curves correspond to different values of $y$. The agreement with QSSEP is very good with $\Delta n^{(2)}=1.3$ obtained as the best fit to the data.
\textbf{c)} Same plot, but collapsing different values of $N$, which nicely converge for increasing $N$. The imbalance $\Delta n^{(2)}=1.4$ for the QSSEP prediction is obtained as a fit to an extrapolation of the data to $N=\infty$.
\textbf{d)} 3d plots of the third cumulant in both models for a cut at $z=0.525$ show a qualitative agreement of the two shapes. 
\textbf{e,f)} Cuts of \textbf{$g^{{\rm Q/A}}_3$} for fixed $z=0.23$ and $y=0.78, 0.23$. The imbalance $\Delta n^{(3)}=-0.77$ here is fitted by hand to the cut with $y=0.78$. The agreement with QSSEP at the given imbalance is much worse for $y=0.23$. A reason could be that here the values of $y$ and $z$ coincide. Notice the \emph{negative} sign for $\Delta n^{(3)}$. \label{fig:mainFig}}
\end{figure*}

\emph{n=1. }We plot $g_{1}^{{\rm A}}$ in Fig.~\ref{fig:mainFig} a.
The local density linearly interpolates
between the right and left boundary, as expected for a diffusive system. 
Fitting the slope, we get $\Delta n^{(1)}=1.71$.

\emph{n=2.} Results for $g_{2}^{{\rm A}}$ are shown in Fig.~\ref{fig:mainFig}~b,c and are compared to $g_{2}^{{\rm Q}}$ for QSSEP from Eq.~(\ref{eq:QSSEPcontinuous-1}). In Fig.~\ref{fig:mainFig} c, the value $\Delta n^{(2)}$ is obtained by extrapolating the numerical data to $N\to\infty$ using
the ansatz $g_{2}^{{\rm A}}=g_{2}^{{\rm Q}}\left(1+\frac{\beta}{\left(\Delta n^{(2)}\right)^{2}N}+\frac{\gamma}{\left(\Delta n^{(2)}\right)^{2}N^{2}}\right)$ and $\beta, \gamma$ additional fitting parameters. We obtain $\Delta n^{(2)}=1.43$ which is slightly below $\Delta n^{(1)}$. Ideally, the two values would agree.

\emph{n=3.} Finally, we show $g_{3}^{{\rm A}}$ in Fig.~\ref{fig:mainFig} d,e,f. The correspondence is
less clear than for the second order but still in good qualitative agreement.
The optimal fit for the density imbalance is $\Delta n^{(3)}=-0.77$
which is way below the first and second order values and maybe more significantly, carries a negative sign. Additionally,
we see significant discrepancies whenever the indices are brought close together, such as in Fig.~\ref{fig:mainFig} f.

\begin{figure}[ht]
    \centering
    \hfill
    \includegraphics[width=0.46\columnwidth]{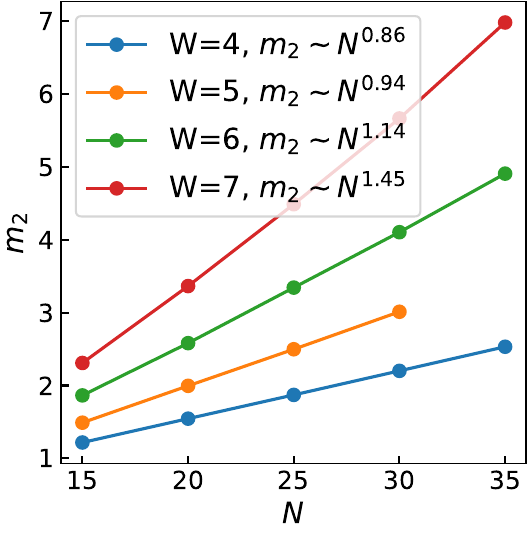}
    \hfill
    \includegraphics[width=0.46\columnwidth]{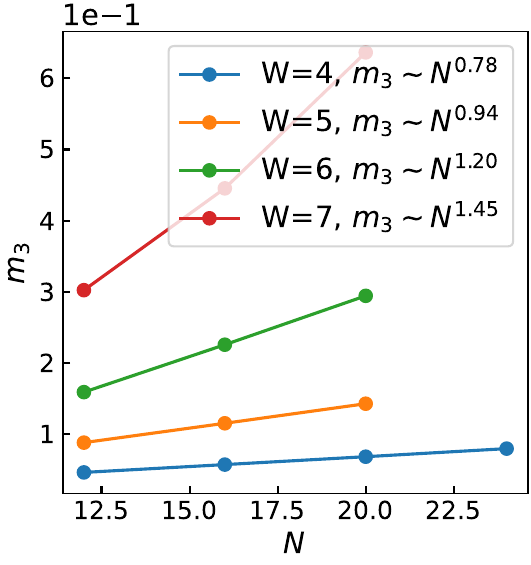}
    \hfill
    \vspace{-2mm}
    \caption{Scaling of the correlations of coherences with system size in the 3d Anderson model. Here $m_2:=\sum_{i_{x}j_{x}}\left[\overline{G_{\boldsymbol{ij}}^{\mathrm{neq}}G_{\boldsymbol{ji}}^{\mathrm{neq}}}^{c}\right]_{\perp}$, $m_3:=\sum_{i_{x}j_{x},k_{x}}\left|\left[\overline{G_{\boldsymbol{ij}}^{\mathrm{neq}}G_{\boldsymbol{jk}}^{\mathrm{neq}}G_{\boldsymbol{ki}}^{\mathrm{neq}}}^{c}\right]_{\perp}\right|$. For values in the interval $W\in[5,6]$, both expressions scale approximately proportional to $N$ as predicted by QSSEP, since cyclic cumulants of order $n$ should scale with $N^{1-n}$.
    \label{fig:Nscaling}}
\end{figure}

Additionally, in Fig.~\ref{fig:Nscaling} we checked for different $W$ if the system size scaling of the second and third cumulants is the one predicted by QSSEP. We find best agreement for $W\in[5,6]$ which is consistent with $W=6$ for which also the shape of $g_n^\text{A}$ matches best with QSSEP  (see Fig.~\ref{fig:mainFig}).

\paragraph{Hints for a proof.}
Here we provide preliminary ideas for a proof of the correspondence. First, note that
the \emph{pure }random dephasing model,
\begin{equation}
dH_{{\rm Deph}}(t)=H_{0}+\sqrt{\gamma}\sum_{i}n_{i}dB_{t}^{i},\label{eq:dephasing}
\end{equation}
where $H_{0}$ is the usual fermionic tight-binding model and $B_{t}^{i}$
are independent Brownian motions at every site, maps \emph{exactly} to the QSSEP when considering the strong dephasing,
long time limit $\gamma\to\infty$, $t\to\infty$, $s:=\frac{t}{\gamma}$
finite, where $s$ plays the role of an effective time \citep{BauerBernardJin_Stoqdissipative}. This model has been
thoroughly studied in the literature, see e.g. \citep{Znidaric__XXdeph,Znidaric_MPSDeph,Znidaric_dephasing,ProsenEssler_Mapping,BauerBernardJin_Stoqdissipative,BastianelloDeNardisDeluca_GHDDeph,dolgirevNonGaussianCorrelationsImprinted2020a,Jin_Quantumresistors},
some of the motivations being its analytical tractability and the
fact that it entails \emph{diffusive }transport in 1d. 
One can suppose that this mapping will still be valid at
finite $\gamma$ but large space and time scales, in some renormalization
group sense that has yet to be made precise. 

Under this assumption,
we are left with understanding how a dynamical noise (\ref{eq:dephasing}) may
generate the same statistical fluctuations as static disordered (\ref{eq:Anderson-1}). One possible route is to examine the diagrammatic
structure of field theories of both models. At the single replica
level, a way to resum the diagrammatic series for the self-energy
is the well-known \emph{self-consistent Born approximation} (SCBA)
\citep{Born1926,AltlandSimonsCondensedmatterfieldtheory}. For the
Anderson model, this approximation is justified for $\left(k_{F}\ell\right)^{d-1}\gg1$,
i.e for $d>1$ and in the mesoscopic limit. Under this assumption,
the imaginary part of the self-energy can be computed in the continuum.
In 3d, it is $\Im\left(\Sigma\right)\approx\left(\frac{t_{z}}{t_{\perp}}\right)\nu\left(E_{F}\right){\cal W}^{2}$ where ${\cal W}:=a^{3/2}W$ and $\nu$ is the density of states.
From the self-energy, one deduces the scattering time $\tau\propto\Im\left(\Sigma\right)^{-1}$
and subsequently the mean free path $\ell\propto v_{F}\tau$ and the
diffusion constant $D\propto v_{F}\tau$.

On the other hand, it turns out that for the dephasing model, the
SCBA is \emph{exact }\citep{dolgirevNonGaussianCorrelationsImprinted2020a,Generictransportformula}\emph{
}which is a direct consequence of the fact that the noise term is
delta correlated in time. The self-energy in $1d$ was computed
to be $\Im\left(\Sigma\right)\propto\gamma$.
Thus, at the one replica level, the two theories match when identifying
\begin{equation}
\gamma\to\left(\frac{t_{z}}{t_{\perp}}\right)\nu\left(E_{F}\right){\cal W}^{2}.   
\end{equation}
Improving this statement would require to look at higher order cumulants
of $G$. For the second cumulant, one
would need to analyze the Bethe-Salpeter equations \citep{LeeRamakrishnanReviewdisorderedelectronicsystems}
and search for a diagrammatic correspondence between the models. We leave this to a subsequent study. 

\paragraph{Conclusion.} We first emphasize some delicate points. The correspondence for the fluctuations of the integrated current $Q_{t}$ between SSEP and diffusive conductors by Derrida et al.\ \citep{RocheDerridaCurrentfluctuationsinSSEP} is valid in a \mbox{\emph{quasi-1d}} geometry. In contrast, the extensive scaling of the quantum mutual information in the Anderson model \cite{GullansHuseEntanglementstructure} is only valid in \emph{3d }. In our study, we found an agreement of the shape of correlations with QSSEP both in 3d and quasi-1d \cite{SM}, but the scaling with system size only matches in 3d. This is to some extend paradoxical, sine the statistics of SSEP is already contained in QSSEP.
Furthermore, the optimal disorder ($W=6$) corresponds to a the mean free path of the order of the lattice spacing \cite{EconomouSoukoulisConductivitydisorderedz}, which suggests that no coarse-graining in the longitudinal direction would be necessary for QSSEP to emerge, at least for cumulants at order 1 and 2.

To sum up, we have shown that, despite fundamental disparities between the Anderson model and the QSSEP, the fluctuations of quantum correlations up to third order in both models are analogous, modulo an adjustment of the imbalance $\Delta n$ at every order. This unexpected correspondence hints at the existence of a universal correlation structure in out-of-equilibrium
quantum diffusive systems. The possibility that free probability plays a central role in this universal structure opens an exciting avenue of exploration. A key question is if the correspondence with QSSEP discussed here can be extended to other models, and ultimately, to formulate a quantum version of the macroscopic fluctuation theory (MFT) \citep{Bernard_QMFT}.
We note that encouraging steps in this endeavor have been made recently with numerical simulations in random unitary circuits \citep{FCSMFTDeNardis}
and experiments in chaotic cold atoms systems \citep{MFTBloch} showing that the classical MFT was able to describe these systems. 


\begin{acknowledgments}
\textit{Acknowledgments.} The authors extend their gratitude to Denis Bernard for past and present collaborations and ongoing encouragement. We are thankful towards Marko Medenjak and Alexios Michailidis for initial collaboration on this subject. We thank Mathias Albert and Christian Miniatura for illuminating discussions on the Anderson model. Part of this collaboration was conducted in the ``Les Gustins'' summer school. 
\end{acknowledgments}

\bibliographystyle{apsrev4-2}
\bibliography{biblio}

\appendix
\newpage
\onecolumngrid


\section{SUPPLEMENTAL MATERIAL}
\section{Results for the quasi-1d Anderson model}
While all results presented in the main text deal with the the $3d$ Anderson model in a cubic geometry (that is $N_x=N_y=N_z$), here we present the results for the quasi-1d case where the longitudinal dimension $N_x\gg N_y,N_z$ is much greater than the transverse dimensions. This is the regime in which the result of Lee et.\ al \cite{LeeLevitovYuPRBUniversalstatistics} is valid, since they assume that the transmission properties of each channel (i.e.\ each perpendicular mode $k_\perp$) are statistically independent. The same assumption also allows for analytic calculations via the DMPK equation \cite{DMPK1,DMPK2,Mello1991Maximum-entropy}. In contrast to this, in a recent numerical study of entanglement in the Anderson model, Gullans and Huse \cite{GullansHuseEntanglementstructure} have shown that the mutual information scales extensively with the volume only in the 3d case, but not in quasi-1d. This is because channels in the 3d case are no longer independent \cite{Jalabert1995Quantum}. Keeping in mind that also in QSSEP the mutual information scales as the volume \cite{HruzaBenardEntanglementQSSEP}, we expected that the quasi-1d regime would be the wrong place to look for a correspondence with QSSEP. Much to our surprise, we find that, besides a wrong scaling with the longitudinal dimension $N_x$ (where $N_y$ and $N_z$ stay fixed), the shape of correlations of coherences matches QSSEP.

\begin{figure}
\centering
\hfill
\subfloat[]{\includegraphics[height=0.27\textwidth]{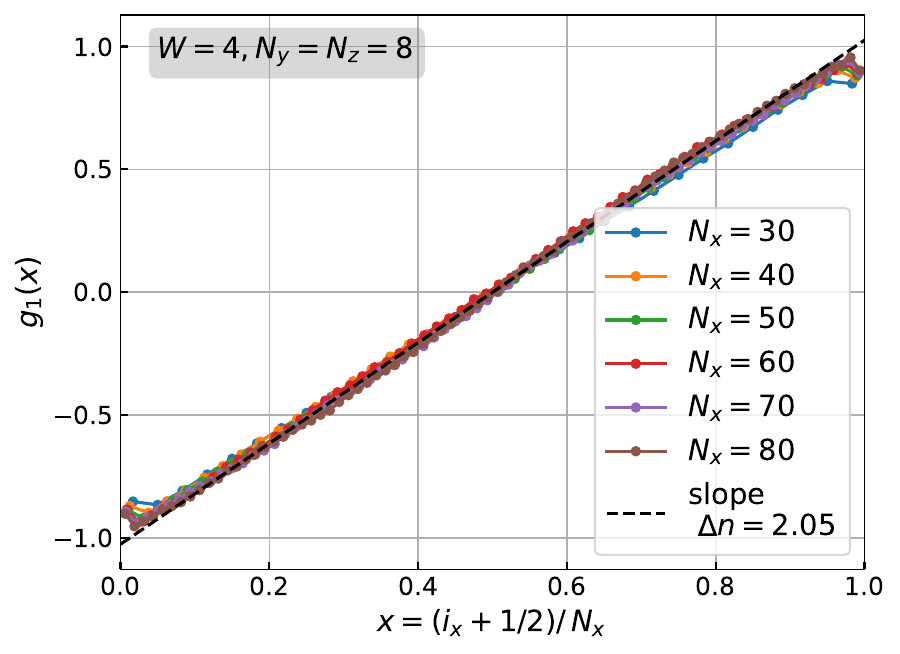}}\hfill
\subfloat[]{\includegraphics[height=0.285\textwidth]{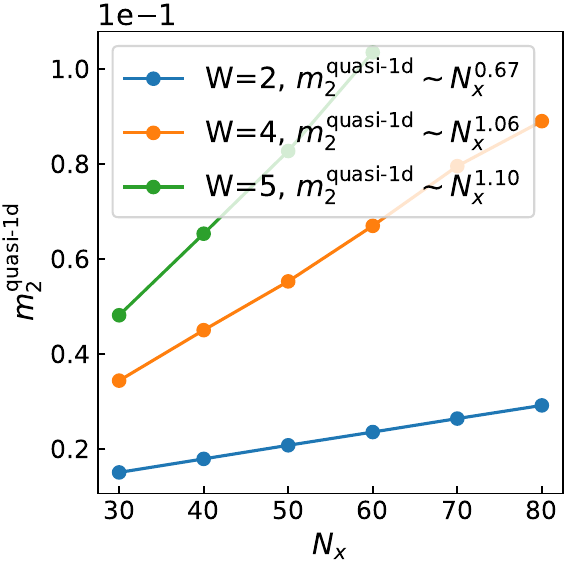}}\hfill
\subfloat[]{\includegraphics[height=0.285\textwidth]{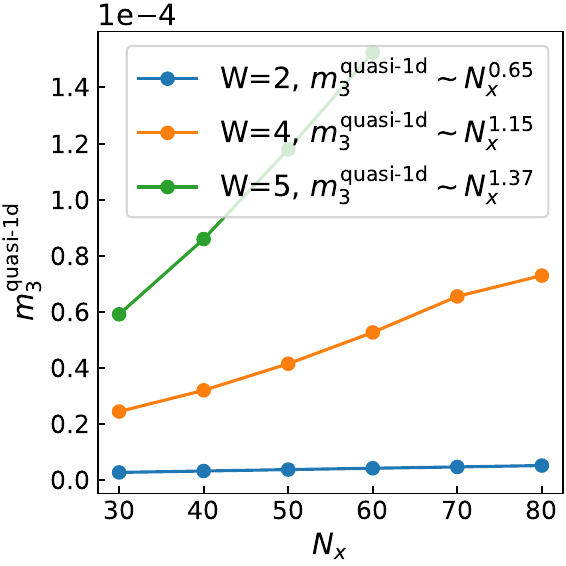}}\hfill
\caption{\label{fig:quasi1d_oder1+scaling} \textbf{(a)} First cumulant (or density) $g_1^\text{quasi-1d}(x)=\left[\overline{G_{ii}^\text{neq}}\right]_\perp$ in the quasi-1d Anderson model for $W=4$ with fixed transverse dimensions $N_y=N_z=8$. In all plots we shift the continuous coordinate $x=i_x/N_x$ by $1/(2N_x)$ to center the data within the interval $[0,1]$ keeping in mind that $i_x=0,\cdots, N_x-1$. \textbf{(b)} and \textbf{(c)} The scaling with $N_x$ assumed in the definition of $g_n^\text{quasi-1d}$ is confirmed for $W\approx 4$ since 
$m_2^\text{quasi-1d}:=\frac{1}{N_x^2}\sum_{i_x,j_x}\left[\overline{G_{\boldsymbol{ij}}^{\mathrm{neq}}G_{\boldsymbol{ji}}^{\mathrm{neq}}}^{c}_{\perp}\right|$ 
and 
$m_3^\text{quasi-1d}:=\frac{1}{N_x^4}\sum_{i_x,j_x,k_x}\left|\left[\overline{G_{\boldsymbol{ij}}^{\mathrm{neq}}G_{\boldsymbol{jk}}^{\mathrm{neq}}G_{\boldsymbol{ki}}^{\mathrm{neq}}}^{c}\right]_{\perp}\right|$ scale approximately linearly with $N_x$.
All data points are obtained as averages over $500$ samples.}
\end{figure}

\begin{figure}
\centering
\subfloat[]{\includegraphics[height=0.26\textwidth]{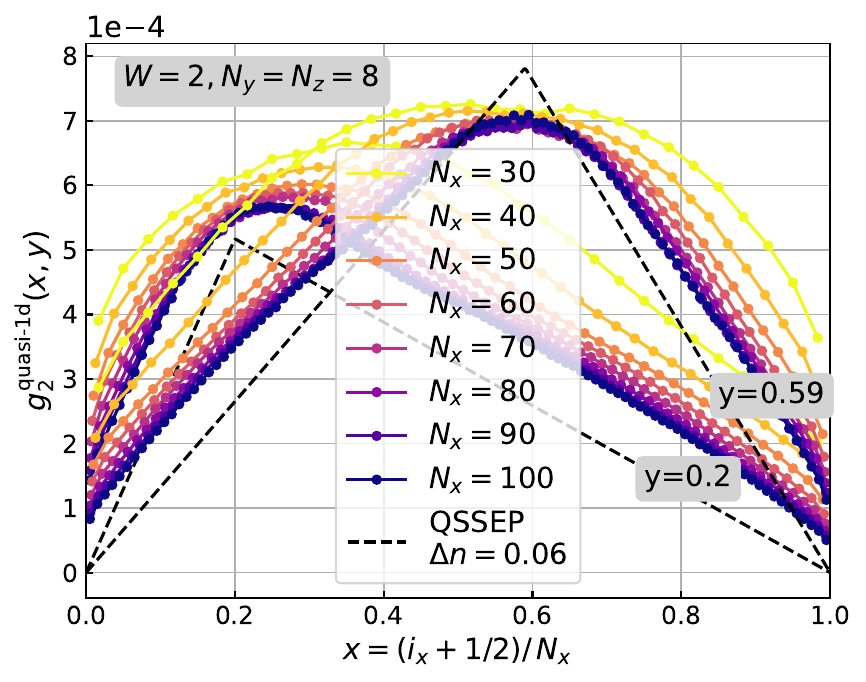}}\hfill
\subfloat[]{\includegraphics[height=0.26\textwidth]{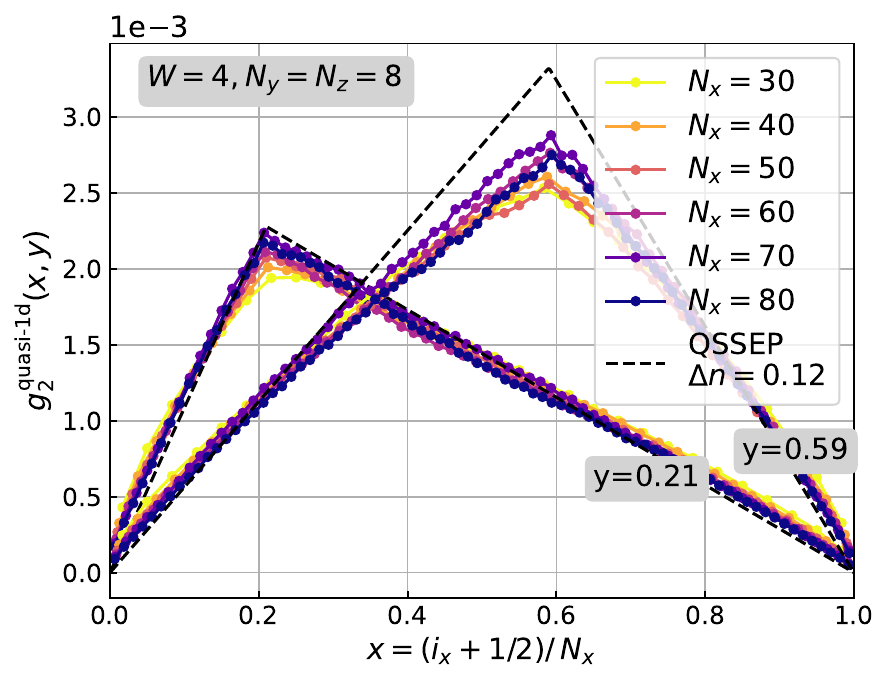}}\hfill
\subfloat[]{\includegraphics[height=0.26\textwidth]{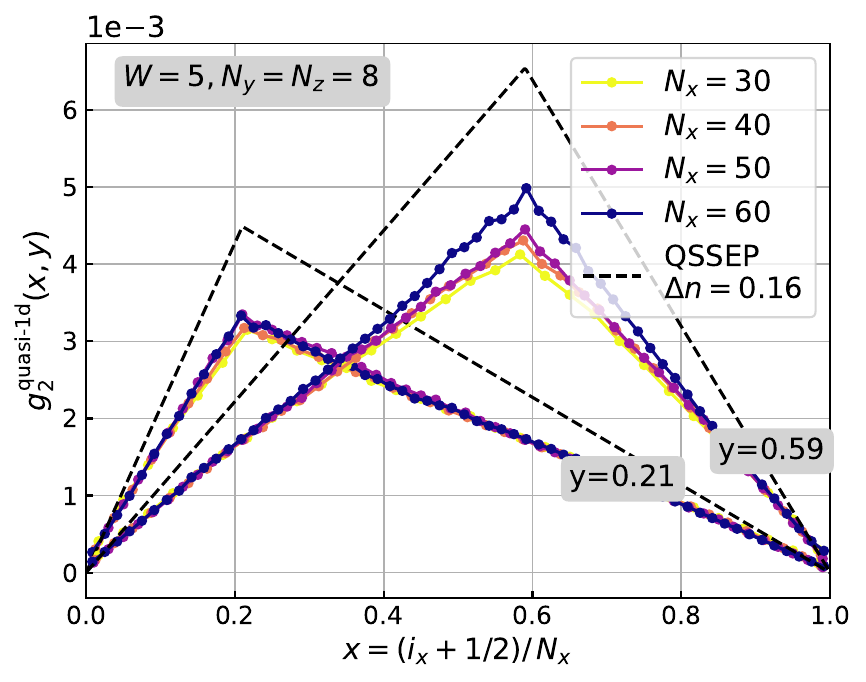}}\hfill
\subfloat[]{\includegraphics[height=0.26\textwidth]{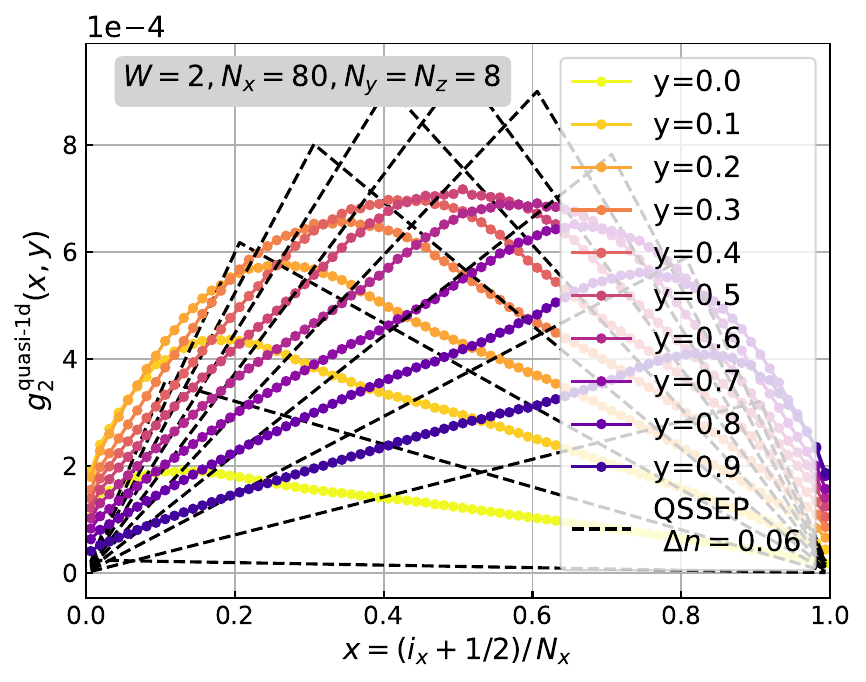}}\hfill
\subfloat[]{\includegraphics[height=0.26\textwidth]{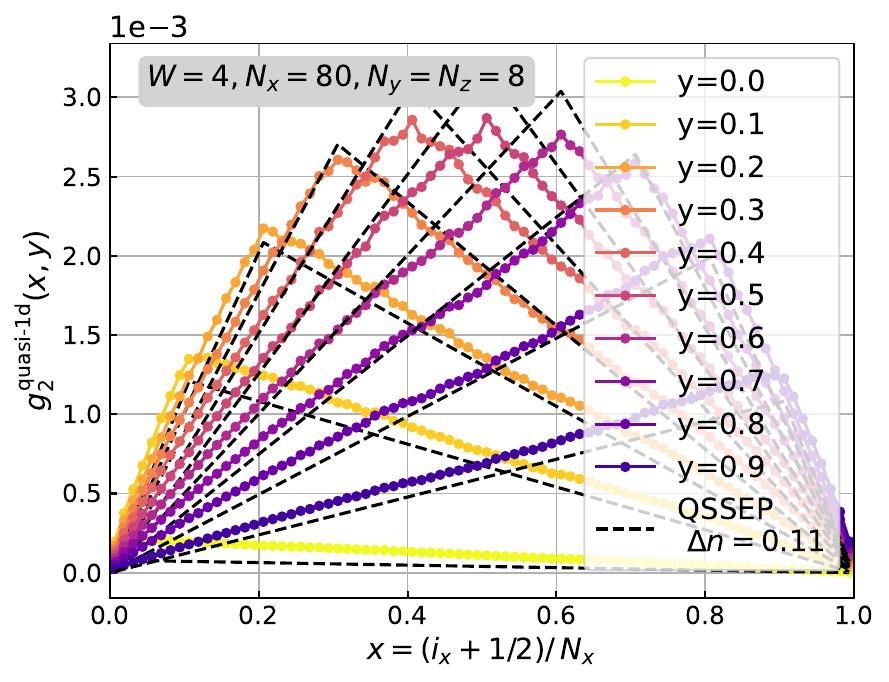}}\hfill
\subfloat[]{\includegraphics[height=0.26\textwidth]{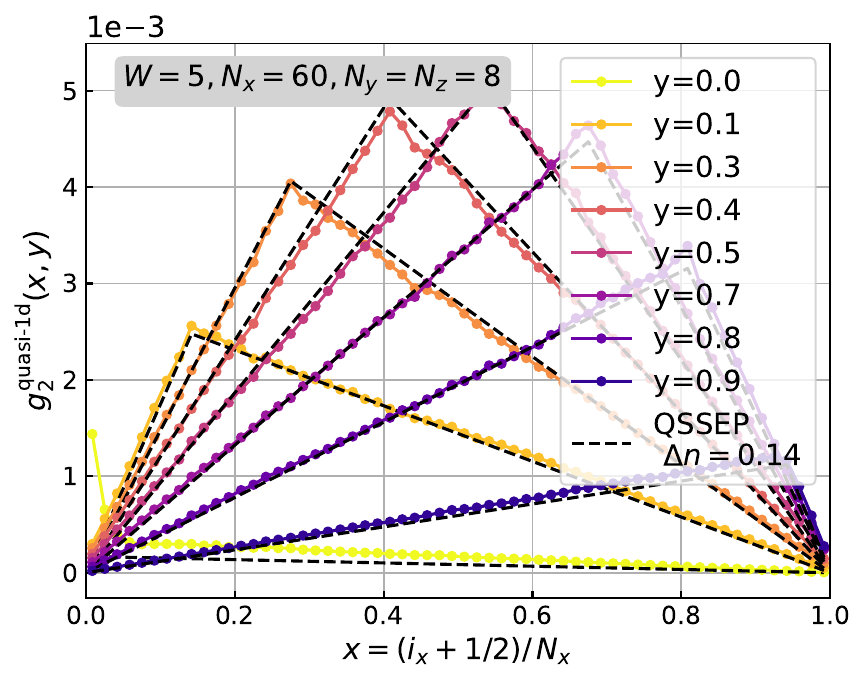}}
\caption{\label{fig:quasi1d_order2}
Second cumulant $g_2^\text{quasi-1d}(x,y)$ in the quasi-1d Anderson model for $W=2,4,5$ and with fixed transverse dimensions $N_y=N_z=8$. In \textbf{(a)}-\textbf{(c)} curves for two fixed values of $y=j_x/N_x=0.21,0.59$ and for different values of the longitudinal dimension $N_x$ are collapsed in order to show that the curves converge for large $N_x$. We observe that convergence is better for lower values of $W$. The value of $\Delta n$ for the QSSEP prediction (dashed line) is found as the best fit to an extrapolation of the data to $N_x\to\infty$. In \textbf{(d)}-\textbf{(f)} curves correspond to a single value of $N_x$ but at different values of $y$. The value of $\Delta n$ for the QSSEP prediction (dashed line) is obtained as the best fit to the data for the corresponding value $N_x$ (and not for $N_x\to\infty$). Note that the correspondence with QSSEP is best for $W=5$. All data points are obtained as averages over $500$ samples.}
\end{figure}

\begin{figure}
\centering
\subfloat[]{\includegraphics[height=0.233\textwidth]{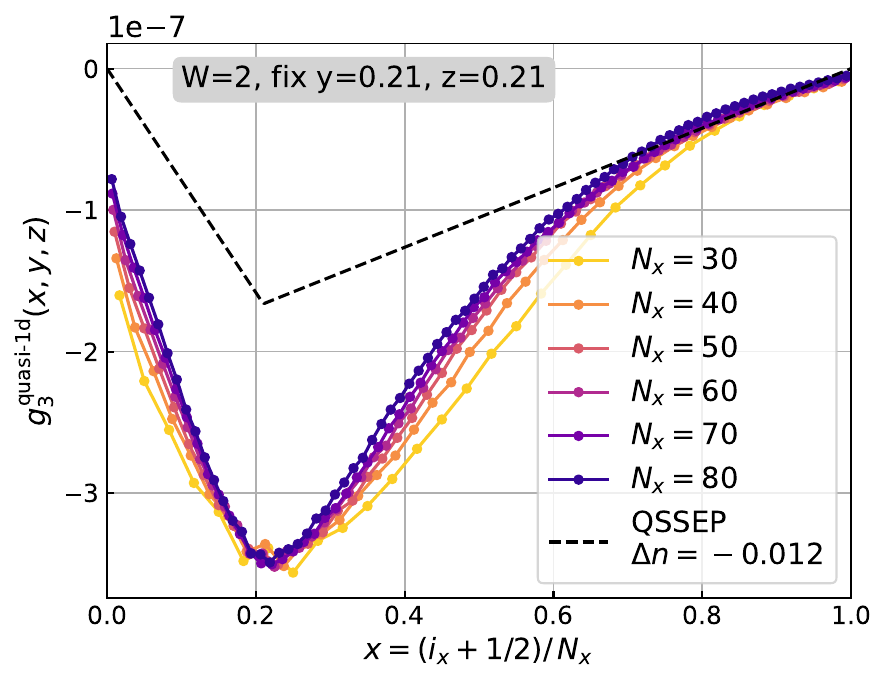}}\hfill
\subfloat[]{\includegraphics[height=0.233\textwidth]{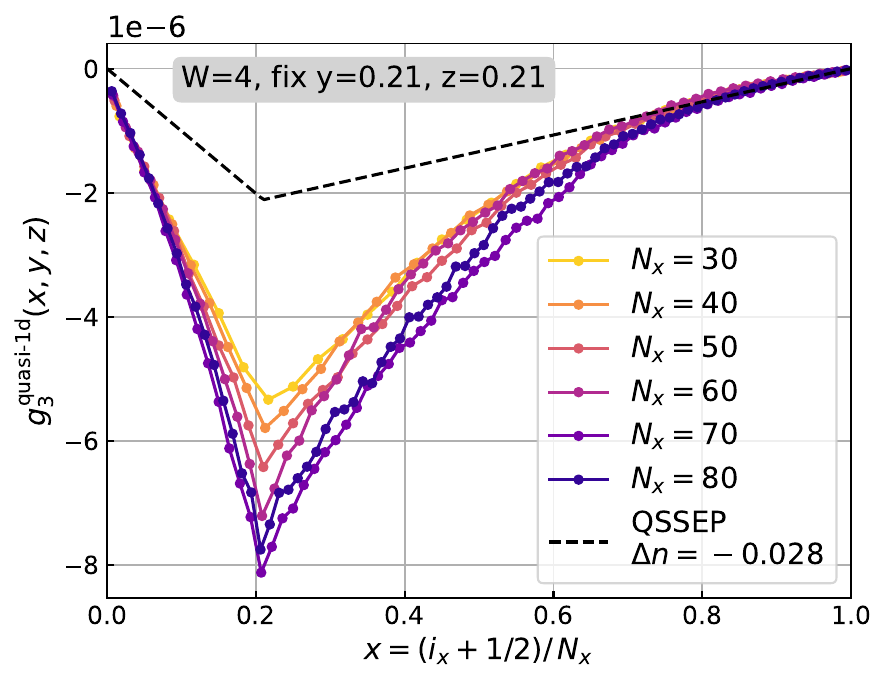}}\hfill
\subfloat[]{\includegraphics[height=0.233\textwidth]{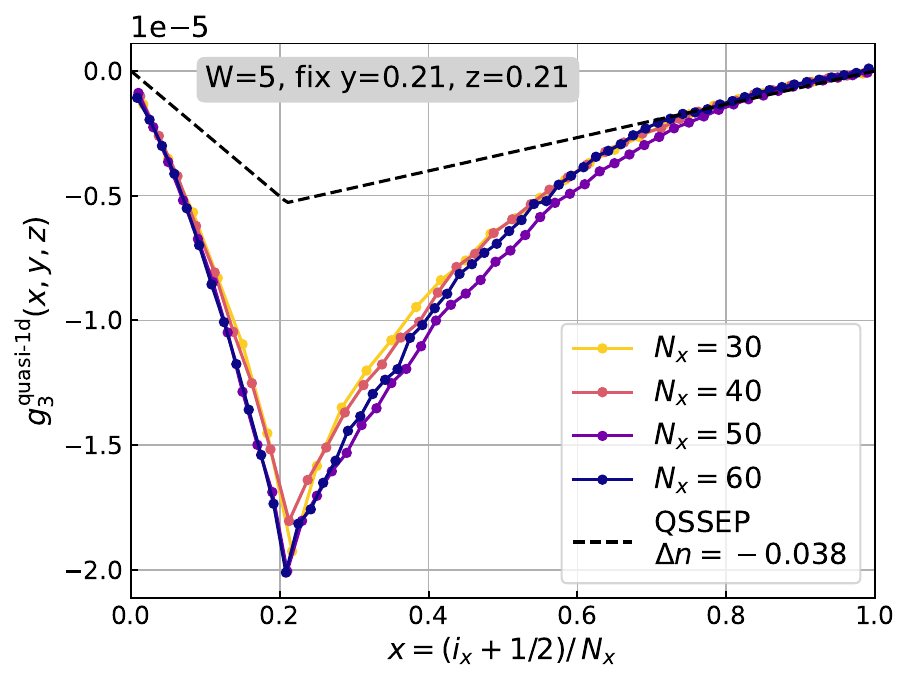}}\hfill
\subfloat[]{\includegraphics[height=0.233\textwidth]{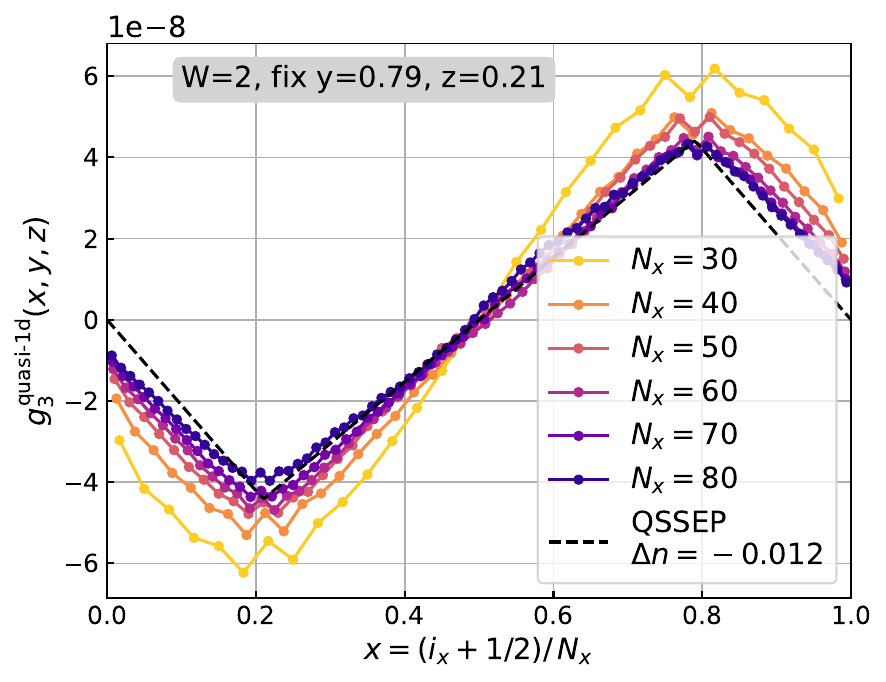}}\hfill
\subfloat[]{\includegraphics[height=0.233\textwidth]{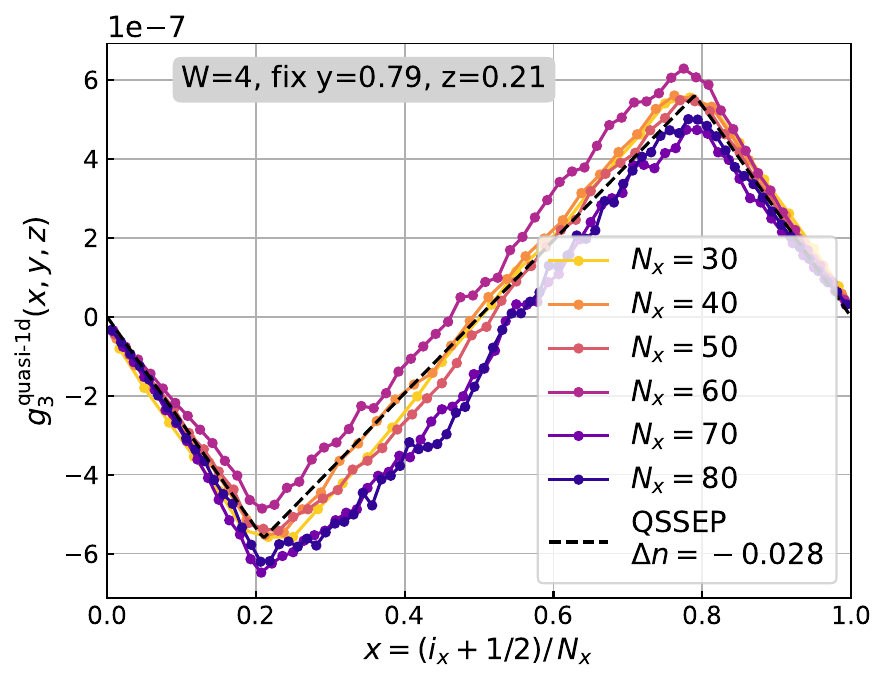}}\hfill
\subfloat[]{\includegraphics[height=0.233\textwidth]{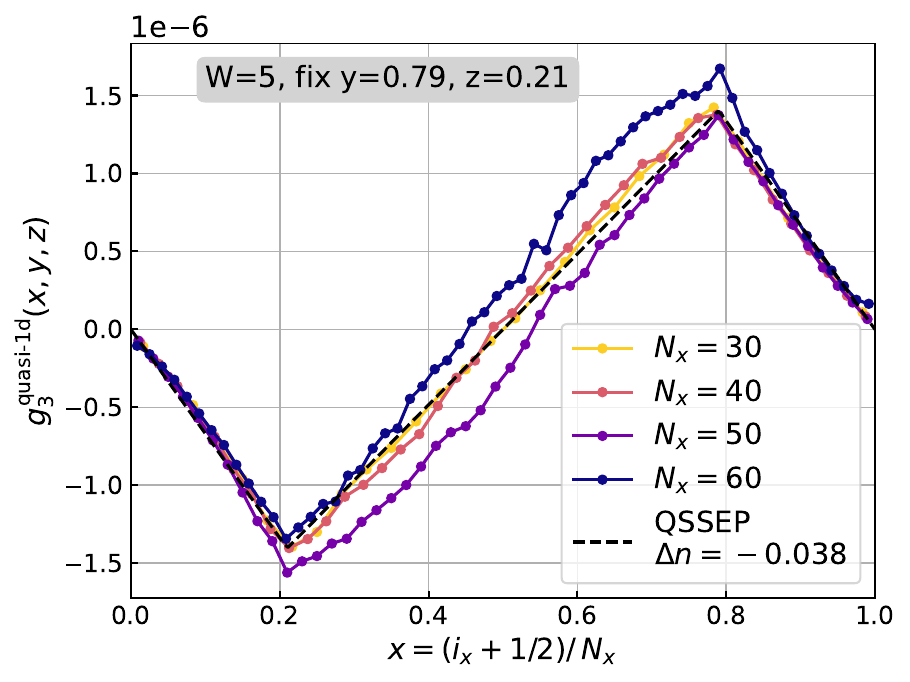}}
\caption{\label{fig:quasi1d_order3} Third cumulant $g_3^\text{quasi-1d}(x,y,z)$ in the quasi-1d Anderson model for $W=2,3,5$ and with fixed transverse dimensions $N_y=Nz=8$. In all plots the value of $z=i_z/N_x=0.21$ is fixed together with $y=j_x/N_x=0.21$ (in \textbf{(a)-(c)}) and $y=0.79$ (in \textbf{(d)-(f)}). The value of $\Delta n$ is fitted by hand to the latter. Note that the convergence with increasing $N_x$ is better for lower values of $W$ which is probably due to numerical errors. Indeed, for $W=5$, $N_y=N_z=8$ and $N_x>60$ our numerical simulation diverges and we don't show these curves here. All data points are obtained as averages over $500$ samples.}
\end{figure}

To show this, we first define the transverse spatial average in analogy to the main text as
\begin{equation}
\left[G_{\boldsymbol{i}_{1}\boldsymbol{i}_{2}}^{{\rm neq}}\cdots G_{\boldsymbol{i}_{n}\boldsymbol{i}_{1}}^{{\rm neq}}\right]_{\perp}:=\frac{1}{(N_y N_z)^n}\sum_{i_{1\perp},\cdots,i_{n\perp}} G_{\boldsymbol{i}_{1}\boldsymbol{i}_{2}}^{{\rm neq}}\cdots G_{\boldsymbol{i}_{n}\boldsymbol{i}_{1}}^{{\rm neq}}.
\end{equation}
with $i_\perp = (i_y,i_z)$ and then define the rescaled cumulants of coherences as
\begin{equation}
    g_{n}^\text{quasi-1d}(x_{1},\cdots,x_{n}):=\frac{1}{N_x^{n-1}}\left[\overline{G_{\boldsymbol{i}_{1}\boldsymbol{i}_{2}}^{{\rm neq}}\cdots G_{\boldsymbol{i}_{n}\boldsymbol{i}_{1}}^{{\rm neq}}}^{c}\right]_{\perp}
\end{equation}
with $x_{k}=i_{k,x}/N_x$ and $\overline{\cdots}^{c}$ the connected disorder average. In contrast to Eq.~\eqref{eq:eq:anderson_cumulants}, here we divide rather than multiply by $N_x^{n-1}$. Fig.~\ref{fig:quasi1d_oder1+scaling} confirms the new scaling with $N_x$ which is best respected for $W\approx4$. Figs.~\ref{fig:quasi1d_order2} and \ref{fig:quasi1d_order3} show the numerical results for the second and third cumulants, respectively. The match with QSSEP is best for $W\approx 5$, though convergence of the curves for different $N_x$ is not as good as for $W=2$. This could be caused by numerical imprecisions which become more important for greater $W$.

Note that all the cumulants of $G^{\rm neq}$ are upper bounded since $|G_{\boldsymbol{i}\boldsymbol{j}}^{\rm{neq}}|\leq1$. Consequently, this scaling with $N_x$ must break down eventually. We expect this to occur when $N_x\gg\xi$ with $\xi$ the localization length, which is the condition to be in the localized regime.


To conclude, we comment on the trace of the $(N_xN_yN_z)\times(N_xN_yN_z)$ matrix $G^{\rm neq}$ and its scaling with $N_x$,
\begin{equation}
    \overline{\text{Tr}({G^{\rm neq}}^{\,n})}^c=\sum_{i_{1x},\cdots,i_{nx}} \left(N_y N_z\right)^n\left[\overline{G_{\boldsymbol{i}_{1}\boldsymbol{i}_{2}}^{{\rm neq}}\cdots G_{\boldsymbol{i}_{n}\boldsymbol{i}_{1}}^{\rm neq}}^{c}\right]_\perp\sim
    \begin{cases}
        N_x^{2n-1}(N_y N_z)^n,  & \text{quasi-1d}\\
        N_x^{2n+1},           & \text{3d with $N_x=N_y=N_z$}.
    \end{cases}
\end{equation}
This suggests, that in terms of scaling with system size, the correlations of coherences in 3d are stronger than in quasi-1d. Why QSSEP still captures the shape of those correlations in both cases remains a riddle to us. 

\newpage
\newpage

\section{Results for varying disorder strength in the 3d case}
\begin{figure}[H]
\centering
\subfloat[]{\includegraphics[height=0.23\textwidth]{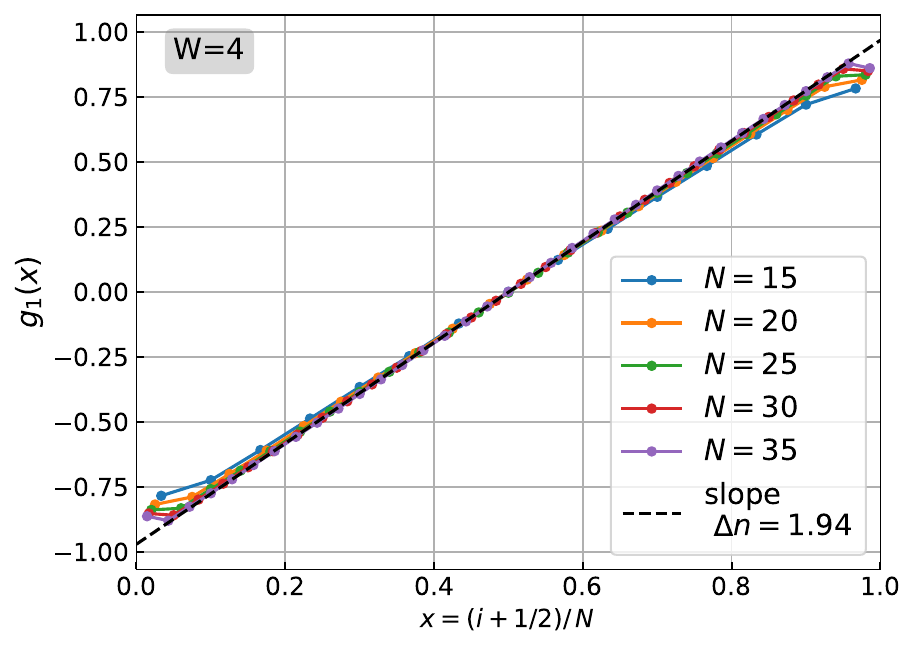}}\hfill
\subfloat[]{\includegraphics[height=0.23\textwidth]{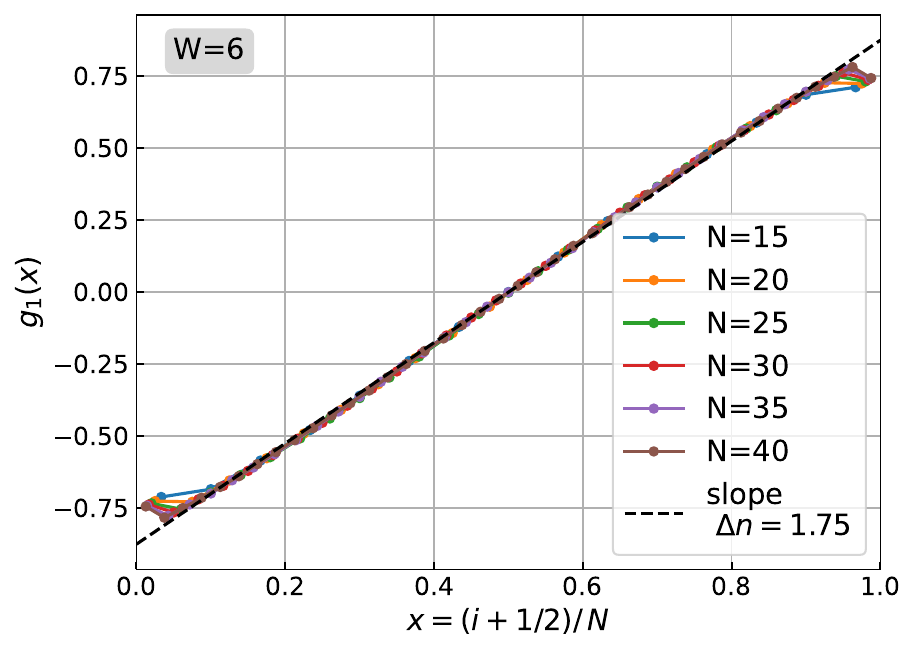}}\hfill
\subfloat[]{\includegraphics[height=0.23\textwidth]{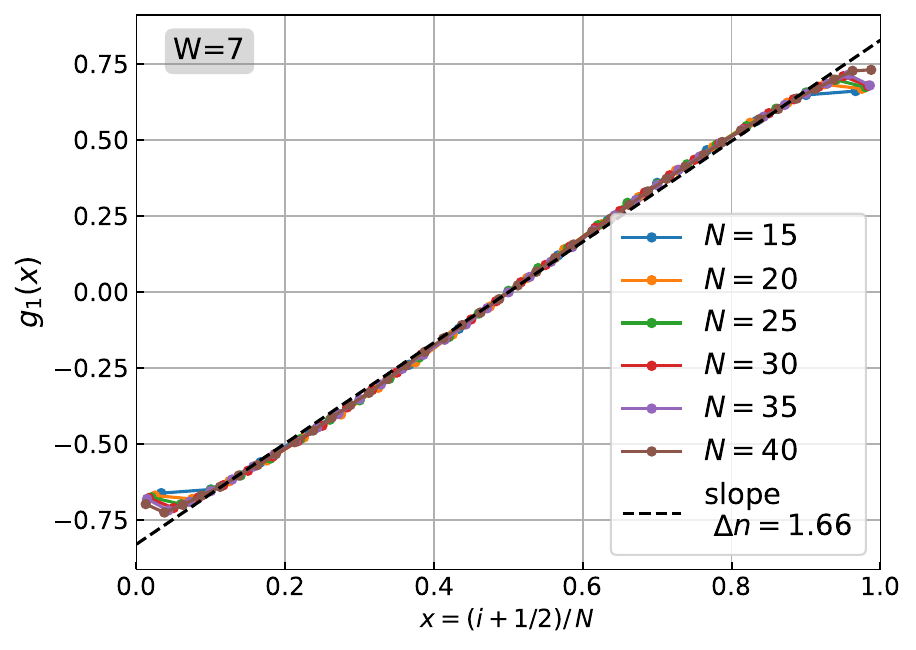}}
\caption{\label{fig:3d_order1}
The density $g_1(x)$ in the 3d Anderson model for $W=4,6,7$. The imbalance $\Delta n^{(1)}$ seems to slightly decrees with increasing disorder $W$. Disorder averages are performed over $1000$ realizations for every point.}
\end{figure}

\begin{figure}[H]
\centering
\subfloat[]{\includegraphics[height=0.23\textwidth]{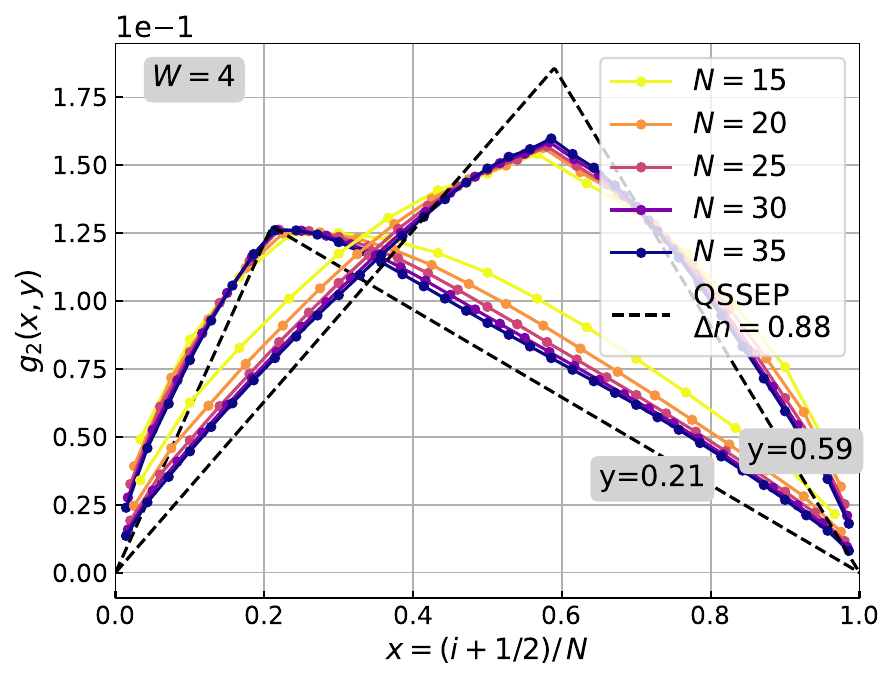}}\hfill
\subfloat[]{\includegraphics[height=0.23\textwidth]{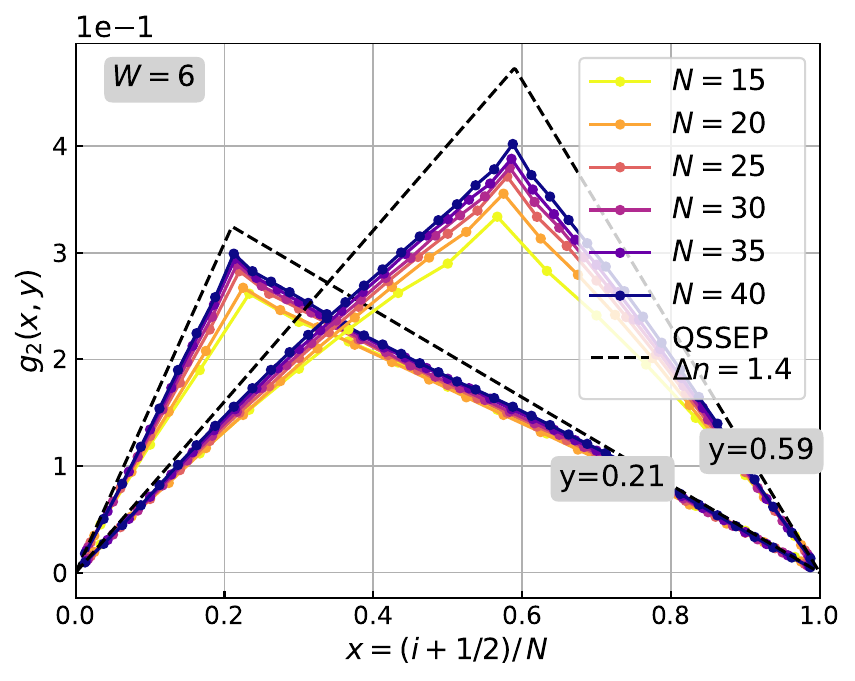}}\hfill
\subfloat[]{\includegraphics[height=0.23\textwidth]{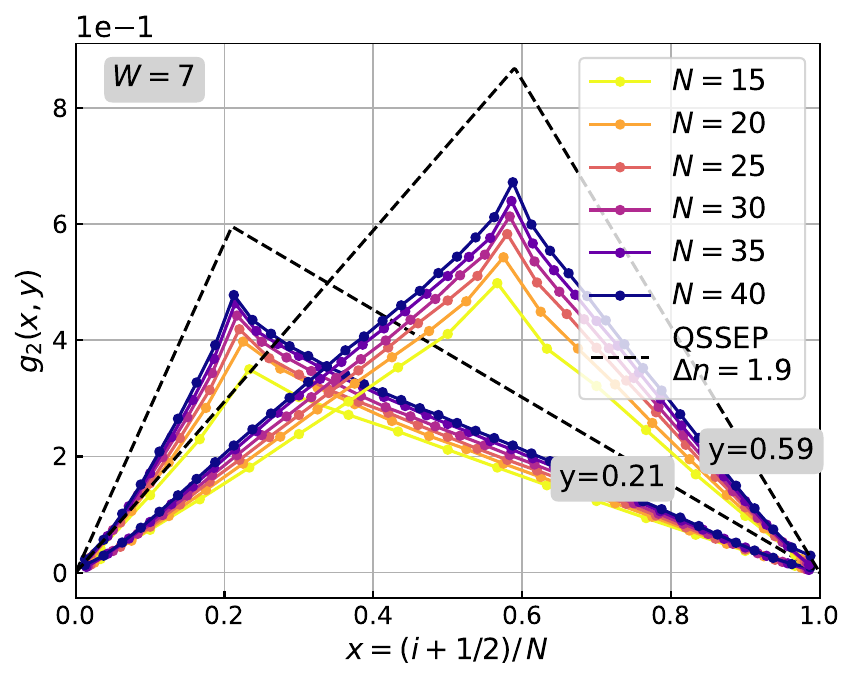}}\hfill
\subfloat[]{\includegraphics[height=0.22\textwidth]{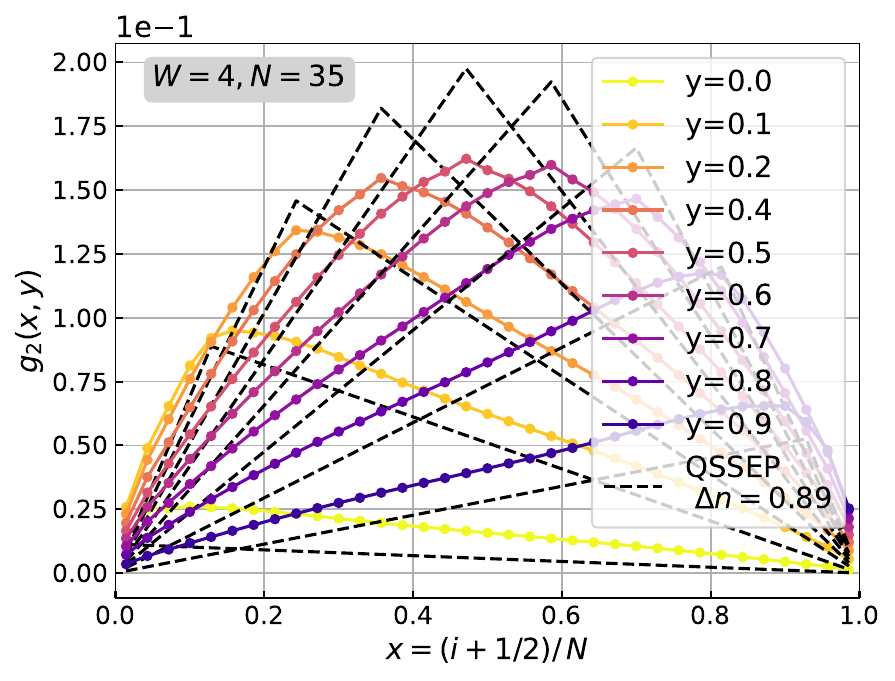}}\hfill
\subfloat[]{\includegraphics[height=0.22\textwidth]{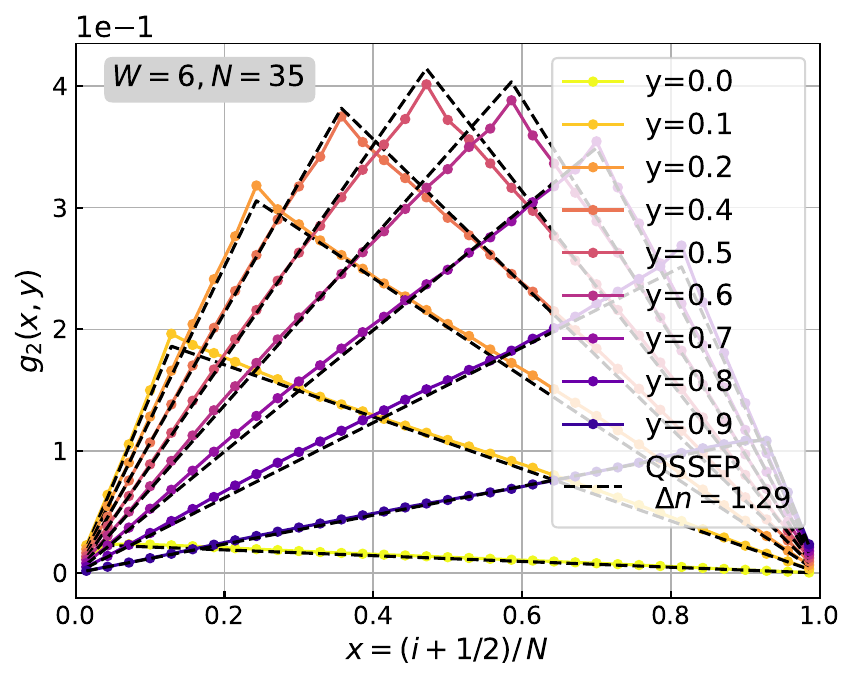}}\hfill
\subfloat[]{\includegraphics[height=0.22\textwidth]{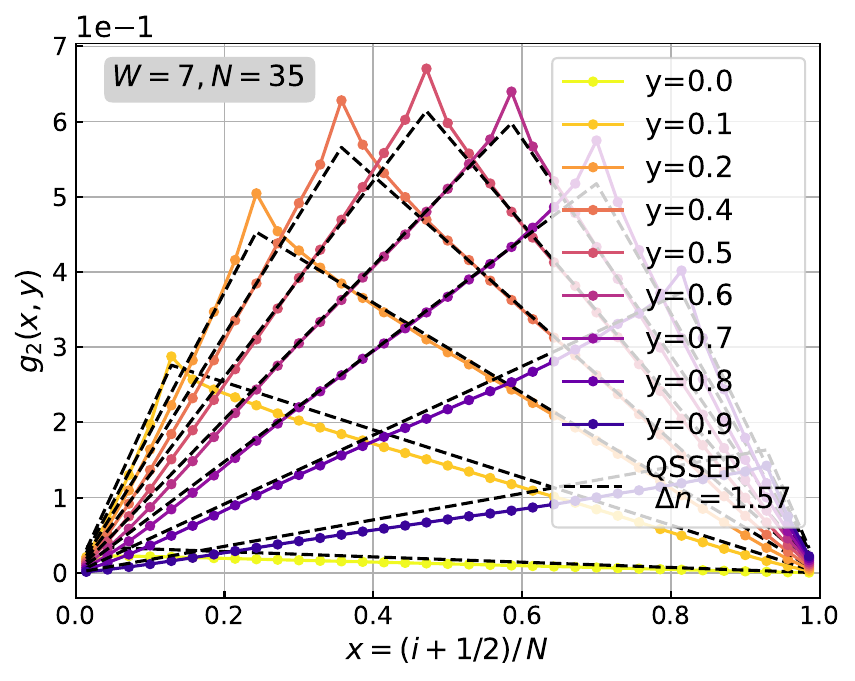}}
\caption{\label{fig:3d_order2}
Second cumulant $g_2(x,y)$ in the 3d Anderson model for $W=4,6,7$. In \textbf{(a)}-\textbf{(c)} curves for two fixed values of $y=j_x/N_x=0.21,0.59$ and for different values of $N$ are collapsed in order to show that the curves converge for large $N$ for all values of $W$. The imbalance $\Delta n^{(2)}$ for the QSSEP prediction (dashed line) is found as the best fit to an extrapolation of the data to $N\to\infty$. In \textbf{(d)}-\textbf{(f)} curves correspond to $N=35$ but at different values of $y$. Here the imbalance for the QSSEP prediction (dashed line) is obtained as the best fit to the data for $N=35$ (and not for $N_x\to\infty$).}
\end{figure}

\begin{figure}[H]
\centering
\subfloat[]{\includegraphics[height=0.22\textwidth]{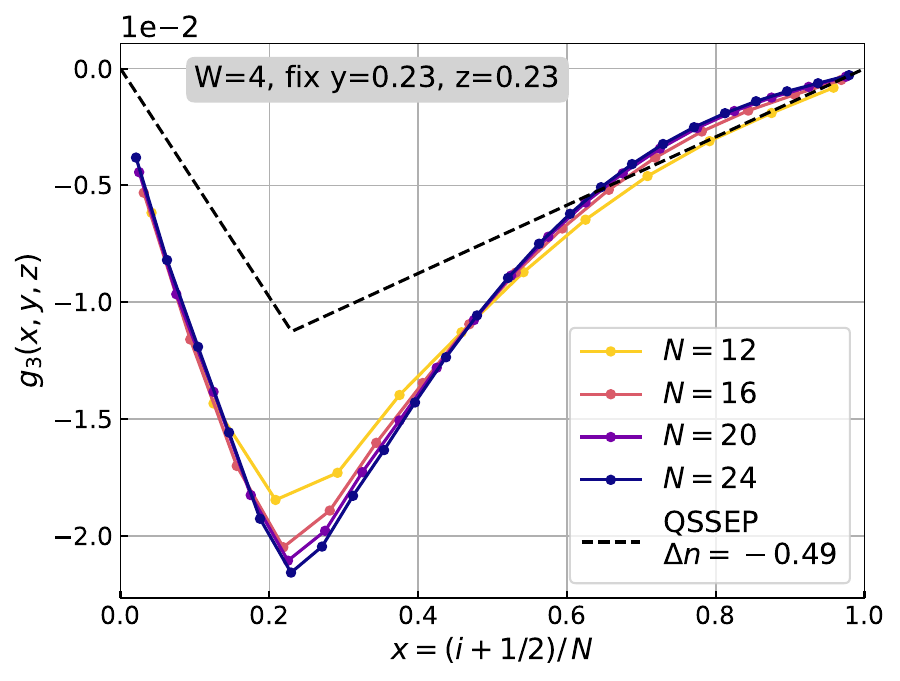}}\hfill
\subfloat[]{\includegraphics[height=0.22\textwidth]{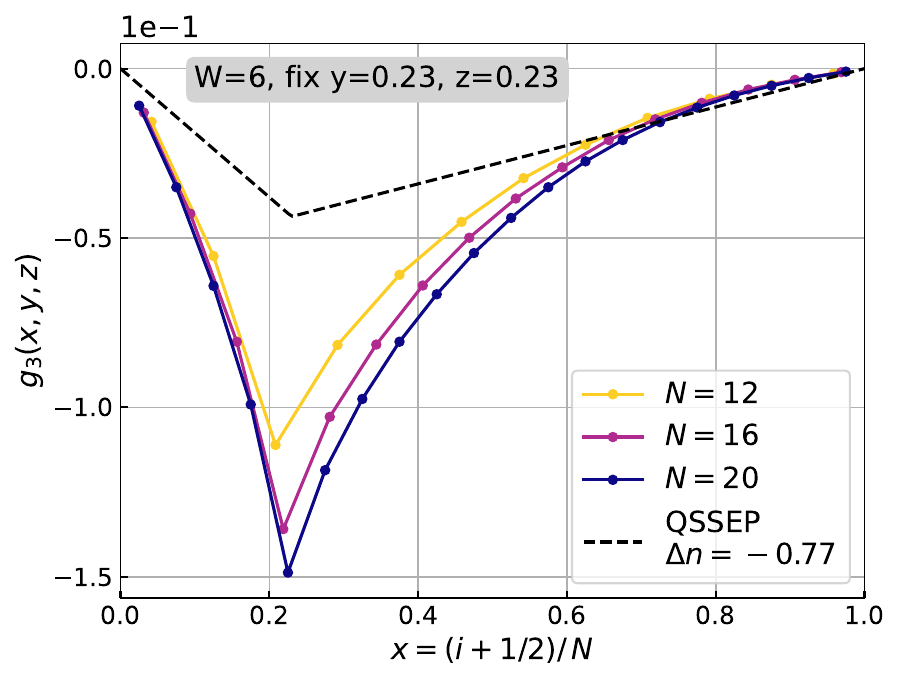}}\hfill
\subfloat[]{\includegraphics[height=0.22\textwidth]{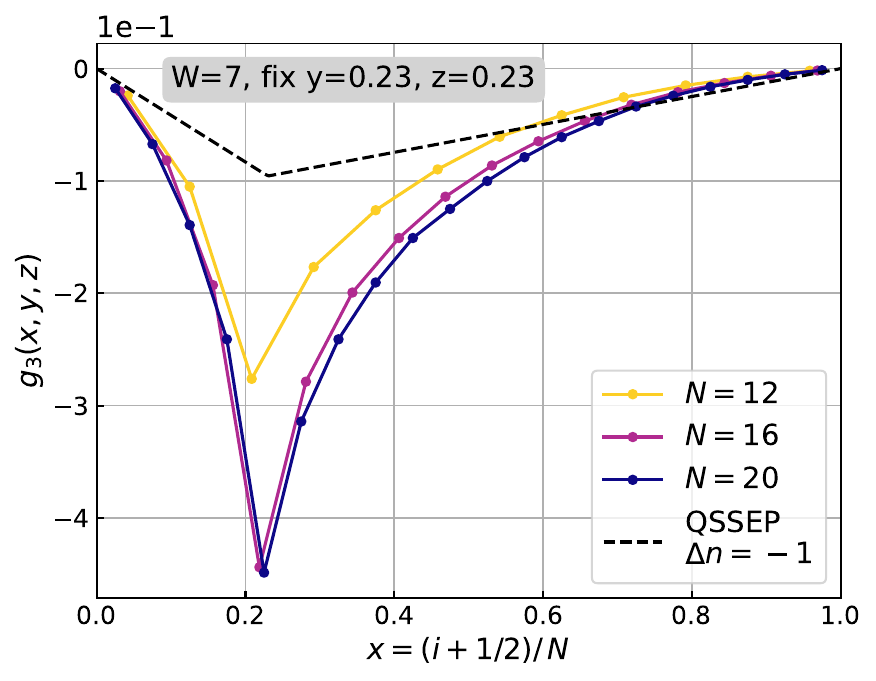}}\hfill
\subfloat[]{\includegraphics[height=0.22\textwidth]{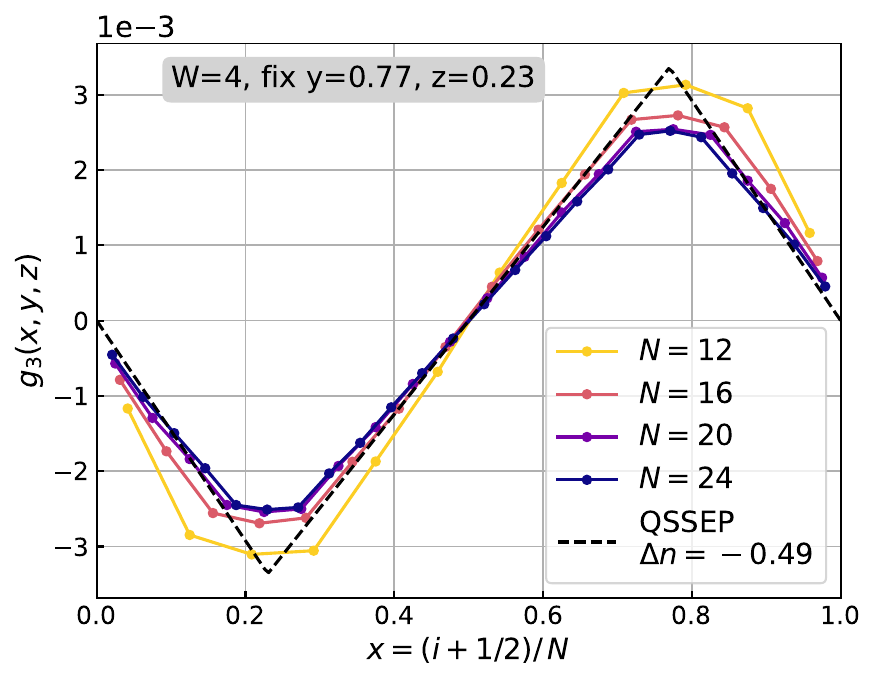}}\hfill
\subfloat[]{\includegraphics[height=0.22\textwidth]{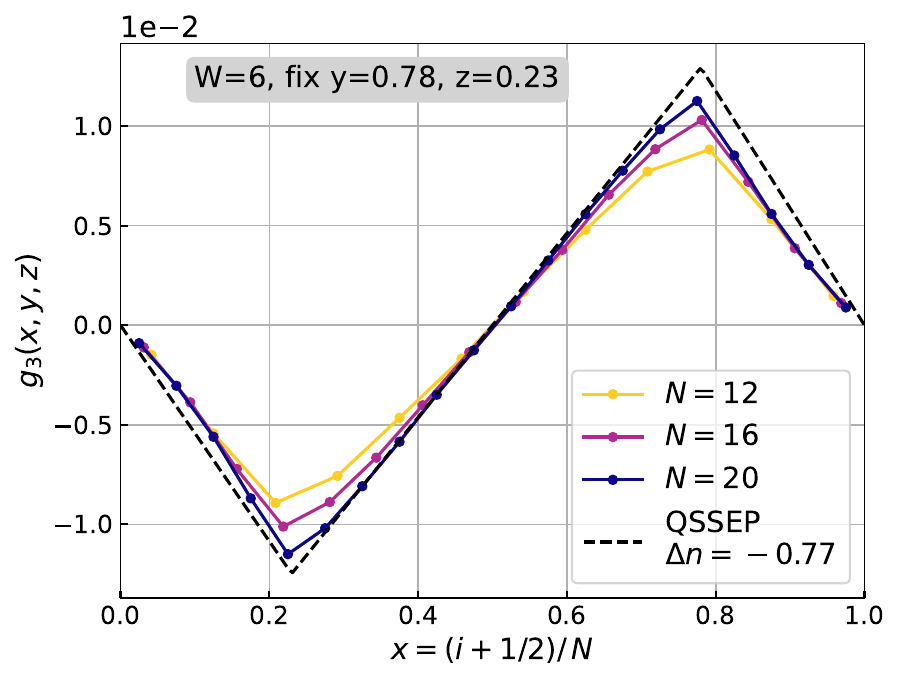}}\hfill
\subfloat[]{\includegraphics[height=0.22\textwidth]{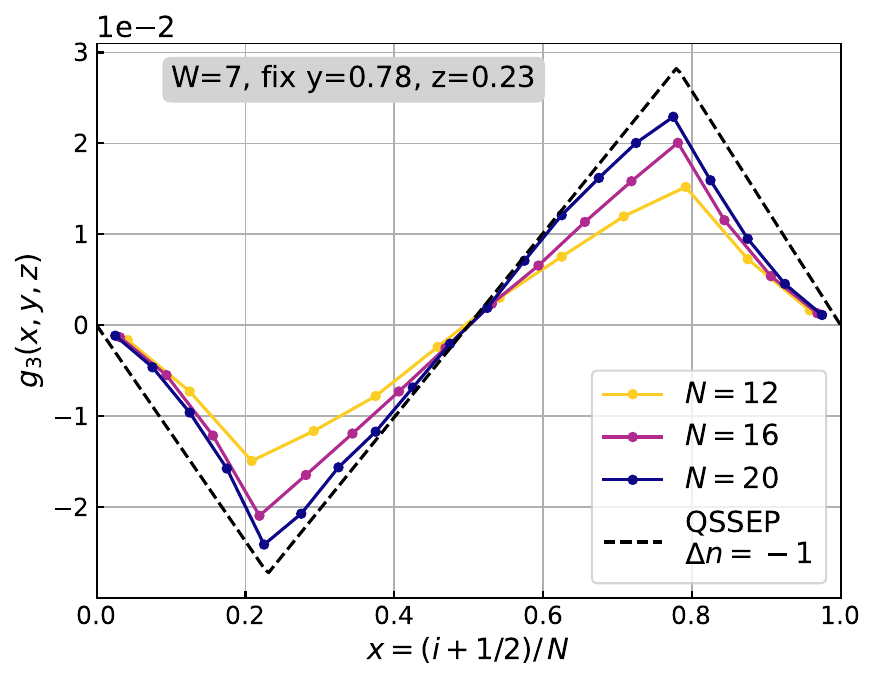}}
\caption{\label{fig:3d_order3}
Third cumulant $g_3(x,y,z)$ in the 3d Anderson model for $W=4,6,7$. In all plots the value of $z=i_z/N_x=0.21$ is fixed together with $y=j_x/N_x=0.21$ (in \textbf{(a)-(c)}) and $y=0.79$ (in \textbf{(d)-(f)}). For the latter, $\Delta n^{(3)}$ is fitted by hand.}
\end{figure}

In the main text we found the best correspondence of correlations $g_{n}^{{\rm Q}}$ in the 3d Anderson model with QSSEP for a disorder strength $W=6$. In Figs.~\ref{fig:3d_order1}, \ref{fig:3d_order2} and \ref{fig:3d_order3}, we show our results for two more values $W=4$ and $W=7$. Note that varying $W$ has an effect on the extracted values of $\Delta n$. Furthermore, we observe that for fixed $W$, the values $\Delta n^{(1)}$ and $\Delta n^{(2)}$ fitted to the first and second cumulant only agree very approximately (best agreement for $W=7$). But $\Delta n^{(3)}$ fitted to the third cumulant differs a lot and caries a negative sign which we don't know how to explain yet.


 \section{Numerical methods for the Anderson model}

For completeness, we recall explicitly the methodology here. We look
for eigenmodes of the system that corresponds to incoming right or
left movers from the reservoirs. A given eigenmode is characterized
by its energy $E$ and \emph{incident} wave vector $\boldsymbol{k}=\left(k_x,k_\perp\right)$ where $k_x$ is fixed by $E$ and $k_\perp$ through the dispersion relation $E_{\boldsymbol{k}}=-\sum_{\nu=x,y,z}2t_{\nu}\cos\left(k_{\nu}a\right)$. Let $\left|\psi_{{\rm \alpha},E,k_\perp}\right\rangle $,
$\alpha={\rm L,R}$ denotes such a single-particle state where ${\rm L/R}$
refers to incoming left/right modes. They fulfill the Schrödinger
equation 
\begin{equation}
H\left|\psi_{{\rm \alpha},E,k_\perp}\right\rangle =E\left|\psi_{{\rm \alpha},E,k_\perp}\right\rangle .
\end{equation}
From this equation, we have a recursive way of relating the wave functions
in a given slice $j_{x}$ to $j_{x}+1$. Explicitly: 

\begin{equation}
\begin{pmatrix}\Psi_{j_{x}+1}\\
\Psi_{j_{x}}
\end{pmatrix}=M_{j_{x}}\begin{pmatrix}\Psi_{j_{x}}\\
\Psi_{j_{x}-1}
\end{pmatrix}.
\end{equation}
where $\Psi_{j_{x}}=\{\psi_{j_{x},j_\perp}\}_{j_\perp}$ is an vector with $N^{2}$ elements specifying the wave function at $\boldsymbol{j}=(j_x,j_\perp)$
and $M_{j_{x}}$ is an $2N^{2}\times2N^{2}$ matrix defined as: 
\begin{equation}
M_{j_{x}}=\begin{pmatrix}\frac{1}{t_{x}}\left({\cal V}_{j_{x}}-E\mathbb{I}-{\cal T}\right) & -\mathbb{I}\\
\mathbb{I} & 0
\end{pmatrix}
\end{equation}
where ${\cal T}$ and ${\cal V}_{j_{x}}$ are $N^{2}\times N^{2}$
matrices whose elements are given by $\sum_{\nu\in\{y,z\}}t_{\nu}\left(\delta_{j_{\nu},j'_{\nu}+1}+\delta_{j_{\nu},j'_{\nu}-1}\right)$
and $\delta_{j_{y}j'_{y}}\delta_{j_{z}j'_{z}}V_{\boldsymbol{j}}$
respectively. In the clean regions, we can always decompose the wave
function in the basis of plane waves:
\begin{equation}
\Psi_{j_{x}\in{\rm L/R}}=\sum_{k_{\perp}}\frac{e^{ik_{\perp}.j_{\perp}}}{N}\left(A_{{\rm L/R},k_{\perp}}e^{\pm ik_{x}j_{x}}+B_{{\rm L/R},k_{\perp}}\mp e^{ik_{x}j_{x}}\right).\label{eq:planewavedecomposition-1}
\end{equation}
where $k_{\perp}:=\left(k_{y},k_{z}\right)$, $j_{\perp}:=\left(j_{y},j_{z}\right)$.
The previous relation to go from plane wave to position basis has
a matrix formulation. For left incomers:
\begin{equation}
\begin{pmatrix}\Psi_{j_{x}+1}\\
\Psi_{j_{x}}
\end{pmatrix}=Q_{{\rm L}}(j_{x})\begin{pmatrix}A_{{\rm L}}\\
B_{{\rm L}}
\end{pmatrix},
\end{equation}
where $A_L=\{A_{{\rm L},k_{\perp}}\}_{k_\perp}$ and $B_L=\{B_{{\rm L},k_{\perp}}\}_{k_\perp}$ are $N^2$ vectors,
\begin{equation}
Q_{{\rm L}}(j_{x})=\begin{pmatrix}U_{{\rm L}} & 0\\
0 & U_{{\rm L}}
\end{pmatrix}\begin{pmatrix}D_{j_{x}+1} & D_{j_{x}+1}^{*}\\
D_{j_{x}} & D_{j_{x}}^{*}
\end{pmatrix}
\end{equation}
where $U_{{\rm L}}$ and $D_{j_{x}}$ are $N^{2}\times N^{2}$ matrices
with elements $U_{{\rm L},j_{\perp},k_{\perp}}=\frac{e^{ik_{\perp}.j_{\perp}}}{N}$
and $D_{j_{x},k_{\perp},k'_{\perp}}=\delta_{k_{\perp},k_{\perp}'}e^{ik_{x}\left(j_{x}+1\right)}$
where we recall that $k_{x}$ depends explicitly on $\left(E,k_{\perp}\right)$ through
the dispersion relation $E_{\boldsymbol{k}}=-\sum_{\nu=x,y,z}2t_{\nu}\cos\left(k_{\nu}a\right)$. 

The transfer matrix relates the amplitudes of the plane waves in the
left reservoir to the ones in the right one: 
\begin{equation}
\begin{pmatrix}B_{{\rm R}}\\
A_{{\rm R}}
\end{pmatrix}=T\begin{pmatrix}A_{{\rm L}}\\
B_{{\rm L}}
\end{pmatrix},
\end{equation}
Since the indices for $A_{{\rm R}}$ and $B_{{\rm R}}$ are swapped
compared to $A_{{\rm L}}$ and $B_{{\rm L}}$, we have in this convention
that $Q_{{\rm R}}=Q_{{\rm L}}$ (see Eq.~\eqref{eq:planewavedecomposition-1})
that we will simply note $Q$ from now on. The explicit expression
of $T$ is 
\begin{equation}\label{eq:transfer_matrix_def}
T=Q(N-1)^{-1}\,M_{N-1}\cdots M_0 \,Q(-1).
\end{equation}
Let $T=:\begin{pmatrix}T_{11} & T_{12}\\
T_{21} & T_{22}
\end{pmatrix}$ where $T_{ij}$ are $N^{2}\times N^{2}$ matrices. For a given right
incomer associated to wave number $k'_{\perp}$ and energy $E$,
we fix $A_{{\rm R},k_{\perp}}:=\frac{1}{\sqrt{2t_{z}\sin\left(k'_{x}\right)}}\delta_{k_{\perp},k'_{\perp}}$
(we use the convention where the incoming current of a given mode
is normalized to $1$), take $A_{{\rm L}}=0$, and we have $B_{{\rm L}}=T_{22}^{-1}A_{{\rm R}}$.
Finally, the elements of $\left|\psi_{{\rm R},E,k_\perp}\right\rangle$ in position space, can be built from the relation: 
\begin{equation}
\begin{pmatrix}\Psi_{j_{x}+1}\\
\Psi_{j_{x}}
\end{pmatrix}=M_{j_{x}}\cdots M_{0}\,Q(-1)\begin{pmatrix}0\\
\left(T_{22}\right)^{-1}A_{{\rm R}}
\end{pmatrix},
\end{equation}
This procedure allows us to obtain the elements of all the single-particle
wave functions $\left|\psi_{{\rm R},E,k_\perp}\right\rangle $
in position space.

For left incomers, the corresponding recursion relation is 
\begin{equation}
\begin{pmatrix}\Psi_{N-1-j_{x}}\\
\Psi_{N-2-j_{x}}
\end{pmatrix}=M_{N-1-j_{x}}^{-1}\cdots M_{N-1}^{-1}\,Q\left(N-1\right)\begin{pmatrix}\left(T^{-1}\right)_{11}^{-1}A_{{\rm L}}\\
0
\end{pmatrix}
\end{equation}
with the explicit expression for $M_{j_{x}}^{-1}$: 
\begin{align}
M_{j_{x}}^{-1} & =\begin{pmatrix}0 & \mathbb{I}\\
-\mathbb{I} & \frac{1}{t_{x}}\left({\cal V}_{j_{x}}-E-\cal T\right)
\end{pmatrix}.
\end{align}
The statistics of the bath is now incorporated by going to second
quantization. Let $a_{\alpha}^{\dagger}(E,k_\perp)$ be
the second quantized fermionic creation operator associated to $\left|\psi_{{\rm \alpha},E,k_\perp}\right\rangle $.
We fix 
\begin{equation}
\langle a_{\alpha}^{\dagger}(E,k_\perp)a_{\alpha'}(E',k'_\perp)\rangle=\delta_{\alpha\alpha'}\delta(E-E')\delta_{k_{\perp},k'_{\perp}}f\left(E,T_{\alpha},\mu_{\alpha}\right).
\end{equation}
And the two-point function $G_{\boldsymbol{i},\boldsymbol{j}}^{{\rm A}}:=\langle c_{\boldsymbol{j}}^{\dagger}c_{\boldsymbol{i}}\rangle$
whose statistics we are interested in is finally given by 
\begin{equation}
G_{\boldsymbol{ij}}^{{\rm A}}=\int dE\sum_{\alpha={\rm L,R}}f\left(T_{\alpha},\mu_{\alpha},E\right)\sum_{k_{\perp}}\psi_{\alpha,E,k_{\perp}}^{*}(\boldsymbol{j})\psi_{\alpha,E,k_{\perp}}(\boldsymbol{i})
\end{equation}
Imposing a small imbalance between the leads $\mu_{{\rm L/R}}=\mp\delta\mu$
and fixing $T_{{\rm L}}=T_{{\rm R}}=0$, we can expand around $E=0$.
Denoting $G_{\boldsymbol{ij}}^{\alpha}(E):=\sum_{k_{\perp}}\psi_{\alpha,E,k_{\perp}}^{*}(\boldsymbol{j})\psi_{\alpha,E,k_{\perp}}(\boldsymbol{i})$ we have 
\begin{equation}
G^{{\rm A}}\approx\delta\mu(G^{{\rm R}}(0^{+})-G^{{\rm L}}(0^{-}))+\int_{-\infty}^{0}dE(G^{{\rm R}}(E)+G^{{\rm L}}(E))
\end{equation}
 and one identifies the non-equilibrium part of $G$ as
\begin{equation}
G^{\mathrm{neq}}:=\delta\mu\left(G^{{\rm R}}(0^{+})-G^{{\rm L}}(0^{-})\right),
\end{equation}
where $0^{\pm}$ means that we can evaluate the energy for any point
in the interval $\left[0,\pm\delta\mu\right]$ but should avoid taking
the exact same point for the left and right movers as this is a special
case. Throughout our simulations we took $0^\pm=\pm0.2$. For convenience, we will avoid dealing with imaginary values
for $k_{x}$ by imposing an anisotropic tight-binding term: $t_{y},t_{z}\ll\frac{1}{2}t_{x}$.

\section{Renormalization procedure}

In this section, we give more details on the renormalization procedure
of the algorithm that was proposed in \citep{Pendrynumerics}. Due to the iterative application of $M_{j_x}$ in the definition of the transfer matrix $T$ in Eq.~\eqref{eq:transfer_matrix_def}, the eigenvalues of the block $T_{22}$ can become very small and can get lost numerically. However, these small eigenvalues dominate of the inverse $T_{22}^{-1}$. Therefore a renormalization procedure is necessary.

First, we introduce the notations $Q^{-1}\left(N-1\right)=\begin{pmatrix}L^{+}\\
L^{-}
\end{pmatrix}$ and $Q\left(-1\right)=(R^{+}|R^{-})$ where $L^{\pm}$ and $R^{\pm}$
are $N\times 2N$ and $2N\times N$ matrices respectively. $T_{22}$
is given by
\begin{equation}
T_{22}=L^{-}M_{N-1}...M_{0}R^{-}
\end{equation}
which can be rewritten as $T_{22}=L^{-}r_{N-1}$ with $r_{n}=M_{n}r_{n-1}$
and $r_{-1}=R^{-}$. The top and bottom half 
\begin{equation}
r_{n}=\begin{pmatrix}r_{1,n}\\
r_{2,n}
\end{pmatrix}
\end{equation}
tend to have large eigenvalues, hence the small eigenvalues relevant for
the inversion of $T_{22}$ will get lost in the iteration. To cure this problem, we define the
$2N\times N$ matrix 
\begin{equation}
r'_{n}:=r_{n}\left(r_{1,n}\right)^{-1}
\end{equation}
such that each block has eigenvalues of order one. Then $\left(T_{22}\right)^{-1}$
is expressed as
\begin{equation}
    \left(T_{22}\right)^{-1}=\left(r_{1,N-1}\right)^{-1}\left(L^{-}r_{N-1}'\right)^{-1}.
\end{equation}

The algorithm to find $r_{N}'$ is as follows:
\begin{itemize}
\item Initialize $r_{-1}'=R^{-}(R_{1}^{-})^{-1}$ 
\item Compute $\tilde{r}_{n}=M_{n}r_{n-1}'$ and $\left(\tilde{r}_{1,n}\right)^{-1}$ 
\item Multiply to get $r'_{n}=\tilde{r}_{n}\left(\tilde{r}_{1,n}\right)^{-1}$ (note the relation $\tilde{r}_n=r_nr^{-1}_{1,n-1}$).
This is the trick of the renormalization algorithm: We obtain $r'_{n}$
without the need to calculate the bad conditioned $r_{n}$. Iterating,
we get $r'_{N-1}$. 
\item From the relation $\left(\tilde{r}_{1,n}\right)^{-1}=r_{1,n-1}\left(r_{1,n}\right)^{-1}$
one deduces $\left(R_{1}^{-}\right)^{-1}\left(\tilde{r}_{1,1}\right)^{-1}\cdots\left(\tilde{r}_{1,N-1}\right)^{-1}=\left(r_{1,N-1}\right)^{-1}$,
from which we get $\left(r_{1,N}\right)^{-1}$.
\end{itemize}

The renormalization procedure for $\left(T^{-1}\right)_{11}$, which
is necessary to treat left incomers, is obtained in a similar way.
\begin{equation}
\left(T^{-1}\right)_{11}=L^{+}M_{0}^{-1}\cdots M_{N-1}^{-1}R^{+}.
\end{equation}
Rename the list $\left[M_{N-1}^{-1},\cdots, M_{0}^{-1}\right]=\left[W_{0},\cdots,W_{N-1}\right]$.
\begin{equation}
\left(T^{-1}\right)_{11}=L^{+}(-1)W_{N-1}\cdots W_{0}R^{+}
\end{equation}
and we are back with the same situation than for $T_{22}$ up to renaming
of the object $M\to W$ and swapping $L^{-}/R^{-}\to L^{+}/R^{+}$. 



\end{document}